\documentclass{article}

\usepackage{graphicx}
\usepackage{color}
\usepackage{xcolor}
\usepackage{geometry}                % See geometry.pdf to learn the layout options. There are lots.
\usepackage{amssymb,amsfonts,amsmath,amsthm}

\usepackage{hyperref}
\usepackage{soul}

\usepackage[round,sort]{natbib}
\newif\ifcomments
\commentstrue
\ifcomments
	\newcommand{\todo}[1]{{\color{blue}\noindent $\bullet$ #1}}
	\newcommand{\clara}[1]{{\color{cyan}{#1}}}
	
	\newcommand{\saray}[1]{{\color{green}{\em #1}}}
	\newcommand{\drt}[1]{{\color{violet}{#1}}}
	\newcommand{\pjm}[1]{{\color{magenta}{#1}}}
\else	
	\newcommand{\todo}[1]{}%{\color{blue}\noindent $\bullet$ #1}}
	\newcommand{\clara}[1]{}%{\color{blue}{#1}}}
	\newcommand{\saray}[1]{}%{\color{green}{\em #1}}}
	\newcommand{\drt}[1]{}%{\color{violet}{#1}}}
	\newcommand{\pjm}[1]{}%{\color{magenta}{#1}}}
	\renewcommand{\st}[1]{}
\fi

\begin{document}

%\linenumbers % just for now

%\title{{Applied community detection: Case studies}}
%\title{An application-driven perspective on community detection in networks: Case studies for community-based scientific research}
%\title{Community detection in networks: Case studies for community-based scientific research}
\title{Case studies in network community detection}

%\author{Saray Shai$^{1,2}$,%\footnote{Corresponding author (sshai@wesleyan.edu)},
%            Natalie Stanley$^{1,3}$,
%            Clara Granell$^{1}$,
%            Dane Taylor$^{1,4}$,
%            Peter J. Mucha$^1$\\
%       $^1$Carolina Center for Interdisciplinary Applied Mathematics, Department of Mathematics, University of North Carolina, Chapel Hill, NC 27599\\
%       $^2$Department of Mathematics and Computer Science, Wesleyan University, Middletown, CT 06459\\
%       $^3$Curriculum in Bioinformatics and Computational Biology, University of North Carolina, Chapel Hill, North Carolina 27599, USA\\
%       $^4$Department of Mathematics, University at Buffalo, Buffalo, NY 14260
%       }
       
\author{Saray Shai, Natalie Stanley,  Clara Granell, Dane Taylor, Peter J. Mucha\\
Carolina Center for Interdisciplinary Applied Mathematics\\ Department of Mathematics, University of North Carolina\\ Chapel Hill, NC 27599-3250}

%\clearpage 
%
%
%\textcolor{blue}{Comment from NS: Something needs to be done about the karate club. The explanation is too overwhelming or maybe should be discussed in another section. Can you just talk about the part that I colored red? }
%
%\drt{DRT: I shortened karate club paragraph. I think the main issue is that we had 5 paragraphs (over a page) on methods. I think this should be shorter, so I summarized these into 2 paragraphs. Please revise :-)} 
%
%\todo{Fig. 1 Do we know of any better visualization showing the nodes numbers that we would like to use here? } \drt{DRT: i think its fine} \saray{saray: I have just created our own visualization in python as Peter suggested. Let me know if it's looking good or not}
%
%\clearpage
\maketitle
\begin{abstract}
%\emph{We'll write a proper abstract when we're done; but for now it's a reminder of our purpose with this $\sim 7000$ word piece. We want to write an application-driven perspective on the uses of community detection. Our description will focus on a set of 6-9 examples from the literature emphasizing in each case (1) the (disciplinary) scientific question, (2) how community detection used to address that question, and (3) what method was chosen and why. Our target audience is networks researchers (in particular, the target audience of the Handbook is graduate students learning SNA) and others with data who might otherwise not immediately consider community detection as a go-to tool. We are hoping that our perspective will provide the reader with a variety of reasons to consider using community detection for their problems and pointers to the methods available to do so.}
%
{Community structure describes the organization of a network into subgraphs that contain a prevalence of edges within each subgraph and relatively few edges across boundaries between subgraphs. 
The development of community-detection methods has occurred across disciplines, with numerous and varied algorithms proposed to find communities.
As we present in this Chapter via several case studies, community detection is not just an ``end game'' unto itself, but rather a step in the analysis of network data which is then useful for furthering research in the disciplinary domain of interest. These case-study examples arise from diverse applications, ranging from social and political science to neuroscience and genetics, and we have chosen them to demonstrate key aspects of community detection and to highlight that community detection, in practice, should be directed by the application at hand.}
\end{abstract}

%\section{Introduction}\label{sec:intro}

Most networks representing real-world systems display community structure, and many visualizations of networks lend themselves naturally to observations about groups of nodes that appear to be more connected to each other than to the rest of the network. One might be reasonably curious about why this is such a common feature across a great variety of real networks, and even more intriguingly, what do the groups mean? 
Considering examples from different disciplines, one can observe that these groups (or communities) often have important roles in the organization of a network. For example, in a social network where nodes represent individuals and edges describe friendships between them, communities can correspond to groups of people with shared interests \citep{Granovetter1973,Zachary1977,mcpherson2001birds,Moody2003}. %%%\pjm{Additional cites here?}\saray{I added some references about homophily, is that ok?}. 
In the graph of the World Wide Web, where a directed edge between web pages represents a hyperlink from one to the other, communities often correspond to webpages with related topics \citep{Flake2000}. In brain networks of interconnected neurons or cortical areas, %\st{or brain regions of the brain and edges account for interconnecting pathways,}
communities can correspond to specialized functional components such as visual and auditory systems \citep{Sporns2016}. In networks representing interactions among proteins, communities can group together proteins that contribute to the same cellular function \citep{Spirin2003}.
Across each of these examples, the communities provide a new level of description of the network, and this intermediate (that is, ``mesoscopic") perspective between the microscopic (nodes) and macroscopic (the whole network) domains proves to be very useful in understanding the essential functionality and organizational principles of a network.

% structural communities and attributes
In particular, one of the motivations to identify communities in many of the aforementioned applications is that the network structure aligns with data attributes
%, for example, that can indicate for each node (i.e.~actor or person) properties
such as age, location, interests, health, race, sex and so on. However, congruent with most community-detection algorithms, we refer to \emph{structural communities} in which there is a prevalence of edges between nodes in the same community versus those between communities. Importantly, this notion is a topological property of the network and is agnostic to attributes. In principle, one can choose other definitions for what constitutes a community, and we note that for attributed (also called annotated) networks there is growing interest in developing community-detection algorithms that utilize both structural and attribute information  \citep{binkiewicz2014covariate,Bothorel2015,newman2016structure,peel2016ground,yang2013community}.
While here we do not explore these possibilities, and focus our attention on communities in the topological sense, it is important to note that there is often positive correlation between community structure and attribute information due to \emph{homophily}  \citep{aral2009distinguishing,mcpherson2001birds}---that is, edges exist preferentially between nodes with similar attributes.  Generally speaking, studying the interplay between attribute information and network structure is complicated due to confounding effects  \citep{shalizi2011homophily}.
%%
%It is, however, not our intention to explore these possibilities here. Instead, we refer to communities in the topological sense, allowing attribute information (if present) to be reflected by model parameters for spreading dynamics.
%% (e.g., a meme may be more likely to spread between persons with similar attributes versus dissimilar attributes). 
%It is important to note, however, that there is often positive correlation between community structure and attribute information due to \emph{homophily}  \citep{aral2009distinguishing,mcpherson2001birds}---that is, edges exist preferentially between nodes with similar attributes.  Generally speaking, studying the interplay between attribute information, network structure and social contagions is complicated due to confounding effects  \citep{shalizi2011homophily}.

Detecting communities in an automated manner is not a simple pursuit, first, because although the qualitative notions of communities may be intuitive, translating such ideas into an appropriate modeling framework can be challenging. In particular, various applications call for different notions of a community, each producing a different mesoscopic description of a network. Second, the computational complexity of community detection can be a fundamental issue; for example, the number of possible partitions of nodes into non-overlapping groups is non-polynomial in the size of the network (and allowing overlapping communities increases the number of possibilities), motivating important work on different heuristics for efficiently identifying communities. Such challenges make community detection one of the most complex---yet fascinating---areas of network science, with a huge and ever increasing number of different algorithms %\st{devised for such purpose that are} 
available in the literature.

{We only indicate a few classes of community-detection methods here, referring the reader to comprehensive community-detection reviews by \citet*{porter2009}; \citet{fortunato2010}; and \citet{fortunato2016} \citep[see also a recent review by][on the conceptual differences between different perspectives on community detection]{Schaub2017}}.
While the ideas of community detection have been around in sociology for decades \citep[see, e.g.,][]{Coleman1964,Moody2003,Freeman2004}, the field has benefited from significant contributions across numerous disciplines, proposing a variety of methods and algorithms for automating community detection.
%One formulation known as 

\emph{Graph partitioning} \citep[e.g.,][]{Kernighan1970,Fiedler1973,Barnes1982,Mahoney2012} %%%\pjm{Add a recent Mahoney reference}\saray{this one?}
spans a large literature across computer science and mathematics, aiming to divide a network into a specified number of groups so that some selected quantity is optimized, such as the number of edges between the groups (i.e., cut size).

\emph{Modularity maximization} \citep{newmangirvan}, a different optimization approach for graph partitioning originating in the physics literature, aims to find the partition with the largest
%A different optimization approach for graph partitioning originating in the physics literature, known as \emph{modularity maximization} \citep{newmangirvan}, maximizes the
difference between the total weight of within-community edges and that expected under a null model---that is, a random-network model with selected properties. Modularity maximization typically leads to more balanced community sizes, can account for degree heterogeneity in the network, and does not require \emph{a priori} specification of the number of communities. However, it is well-known to suffer from a resolution limit \citep{fortunato2007resolution}, and it is not at all clear how to best interpret the different numbers of communities that can be obtained by varying resolution parameters \citep{Reichardt_Bornholdt_2006,njp08}.

%One class of methodologies stemming from statistics is 
{\emph{Statistical inference} \citep[e.g.,][]{Hastings2006,Ball2011,degreeCorrect,Peixoto2013,Peixoto2014}, arising from the statistics literature, typically aims to identify a parametrized generative model that describes the network (e.g., with maximum likelihood). For example, {stochastic block models} \citep{Fienberg1981,Holland1983,Snijders1997} assume {for a} given partition {that} the {edge} probability {between} nodes depends on their community {memberships} \citep[see more details in a recent note by][on the current developments in community detection in the context of stochastic block models]{Abbe2017}.}

{Cut size, modularity, and likelihood all define objective functions that measure the ``goodness'' of the partitions (or, in some cases, sets of communities that may or may not cover the network) and are generally NP-hard optimization problems. In particular, in most cases finding the conclusively best community assignment is effectively computationally equivalent to checking a non-vanishing fraction of all possibilities, which grows exponentially with system size. %While the time required to obtain the definitive exact solution is infeasible for large networks,
Fortunately, many algorithms have been developed to efficiently provide approximate solutions, including %\st{perturbative (i.e., variational)} 
a variety of iterative \citep{Blondel2008,Kernighan1970,Peixoto2014} and spectral \citep{Fiedler1973,Barnes1982,Newman2006} methods.}

{At the same time, numerous heuristics have been developed for community detection that do not necessarily optimize a global objective function but nonetheless have proven to be useful. These often fall into two categories: agglomerative methods which are akin to hierarchical clustering \citep{Hastie2001}; and divisive methods, such as iteratively partitioning a network by some local measure \citep[such as edge betweenness,][]{Girvan11062002}.}

{A number of other community-detection methods stem from analyses of \emph{dynamical systems} on a network, including the Potts model for spin systems \citep{Wu1982,Reichardt2004}, random walks \citep{Zhou2003,Pons2005,rosvall2008maps,Jeub2015,delvenne2010stability}, and oscillator synchronization \citep{arenas2006synchronization,Li2008}. Such approaches are directly applicable for studying these respective dynamical systems and in some cases are closely related or even equivalent to one of the above-mentioned quality functions \citep{Fiedler1973,rosvall2008maps,delvenne2010stability}.}
Conversely, community structure can have a profound effect on dynamics taking place on networks [e.g., the spread of information across social networks \citep{aral2009distinguishing,mcpherson2001birds,melnik2014dynamics,o2015mathematical,ugander2012structural,weng2013virality}, random walks and heat flow \citep{delvenne2010stability,mucha2010community}, cascades \citep{galstyan2007cascading,gleeson2008cascades}, and synchronization \citep{arenas2006synchronization,skardal2012hierarchical}] and adopting a community-based perspective provides a useful vantage point to study these dynamics.

{These are just a small sample of the many community-detection methods that have been developed, and we in no way intend this Chapter to be a comprehensive review of all methods. Rather, here we present examples from different scientific disciplines demonstrating the useful application of community detection. In particular, we aim to emphasize community detection as a tool {for studying networks}. Identifying communities is often just a first step in data analysis as it opens up many possibilities {for further} study. {We illustrate this idea with a well-known example shown in Fig.~\ref{fig:karate}, the Zachary Karate Club \citep{Zachary1977}.}

\begin{figure}[h]
\begin{center}
\includegraphics[trim={0cm 3.2cm 0cm 2.25cm},clip,scale=0.7]{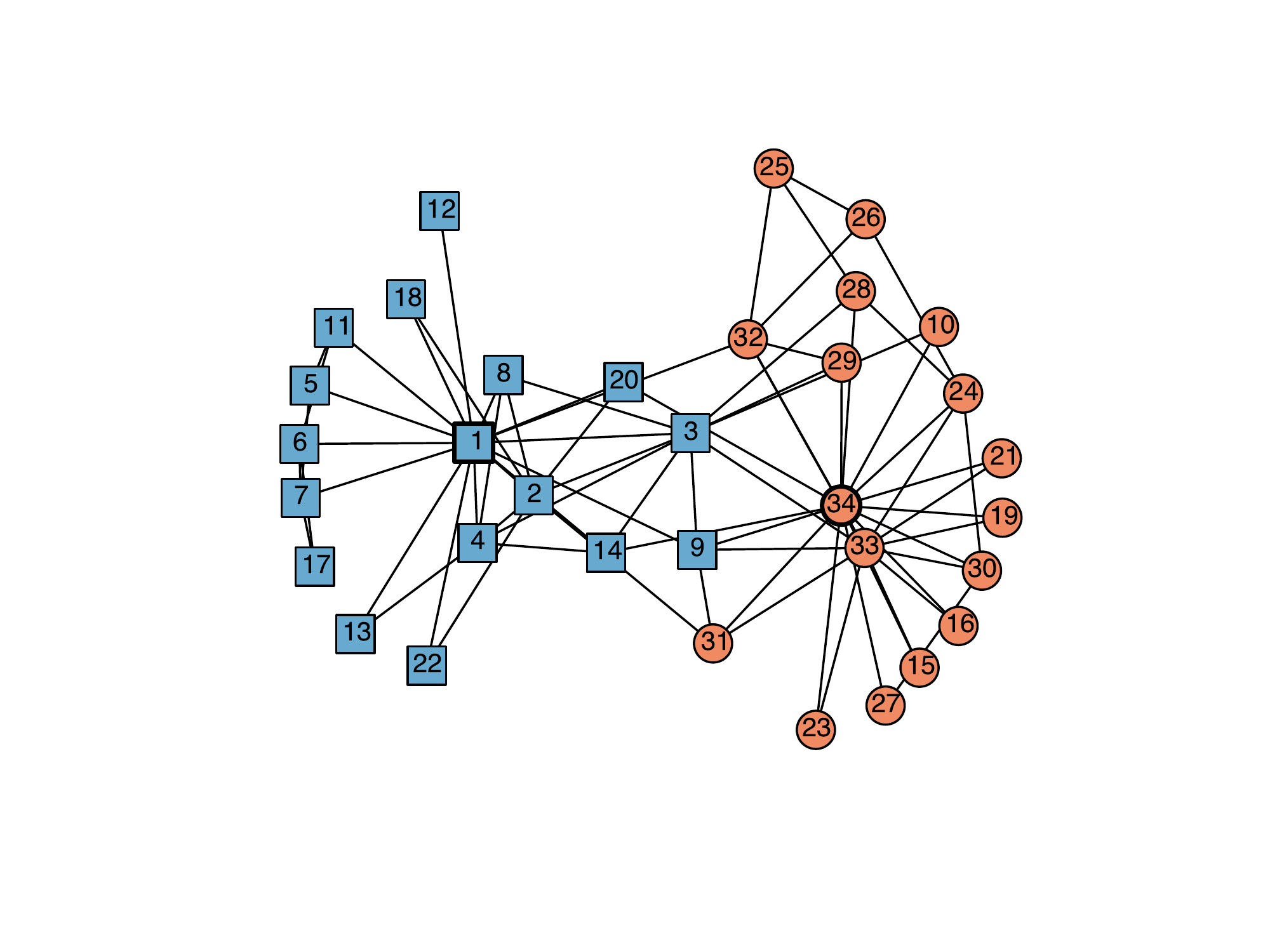}
\caption{Zachary Karate Club Network \citep{Zachary1977}. Node colors and shapes indicate the club division that occurred, with the instructor (node 1) and the president (node 34) shown in bold.
%This spring-directed layout visualization was generated in Python using the NetworkX library.
}
\label{fig:karate}
\end{center}
\vspace{-.5 cm}
\end{figure}

{The Karate Club network developed by Zachary~\citeyearpar{Zachary1977}, through observing the interactions between members of a club during the two-year period 1970--1972, represents the friendships between 34 of the club members as an aggregated, weighted network.
%{The Zachary Karate Club network was developed by Wayne W. Zachary~\citeyearpar{Zachary1977} by observing the interactions between members of a club during the two year period 1970--1972, which he represented by an aggregated weighted network encoding the friendships between 34 of the club members.
During this period, there was a club division (indicated by node colors and shapes in Fig.~\ref{fig:karate}) due to a conflict between the club instructor and the president (nodes 1 and 34, respectively)}. %\st{Due in part to having access to the real division of the dataset in two groups,}
Due in part to this ``ground truth'' division and the network's small size and simple structure, the Zachary Karate Club has become a common example for demonstrating community-detection algorithms.
%Indeed, despite the apparent simplicity of the problem, still most community detection algorithms do not provide the expected partition of the network.
%\st{(Indeed, it's use---and over use---has even inspired the ``Zachary Karate Club Club" of networks scientists using the data as an example} \cite{clubclub}.)
% since one expects the detected communities to align with the actual memberships of the two smaller clubs.
%This network, known as the Zachary Karate Club network, was studied by Wayne W. Zachary for a period of three years from 1970 to 1972~\cite{Zachary1977}. At some point during the observation period, a conflict between the club president and the instructor has led to the split of the club into two smaller clubs (shown in white and pink) supporting the instructor and the president, respectively. to what new club will each member (except member \#9) join
%Zachary predicted the division assuming that members would choose preferentially to be in the subgroup where most of their friends were.
%in the club with their friends. 
Zachary demonstrated that most of the members chose to be in the subgroup best associated with their friends. Specifically, his use of a cut algorithm to define a split of the network into two subgroups almost perfectly reproduced the real-life split of all but one of the members.
%\st{Interestingly, his prediction was correct}. 
Node 9 didn't choose to join the president's new club despite the larger number of ties that linked them, apparently because he 
%\st{The reason for that is that node 9} 
was only three weeks away from completing a four-year quest for a black belt, requiring his allegiance to the instructor.
% \citep{Zachary1977}.

%who
%%Yet, why did Zachary fail to predict the decision of node \#9? 
%%It turns out that at the time of the actual split, the person corresponding to node \#9 
%was three weeks away from completing a four-year quest {for} a black belt, {requiring his allegiance to} 
%%which he could only do with 
%the instructor. 
%\textcolor{red}{ This is obviously something that could not be captured by network structure \drt{alone}. In other words, in this case the correct assignment of nodes to communities, also known as the \textit{ground truth} is not represented in the data, and therefore should not be the objective of a community detection algorithm~\cite{peel2016ground}. Thus, the definition of what constitutes a community also depends on how patterns in the data translate into structural properties.}

{This seemingly odd behavior of node 9 highlights three important lessons:
(I)~adopting a community-based approach to network analysis provides a vantage point to ask new research questions; 
(II)~one must be cautious when comparing the output of a community-detection algorithm to known information on the network (frequently referred to as ``ground truth''), as the latter might include important additional information not captured by the network topology \citep{peel2016ground}; and finally
%%%(III)~[they] predict the virality of memes based on early spreading patterns in terms of community structure. We now describe further each of these results.
(III)~applied community detection should incorporate domain knowledge to choose appropriate methodologies, develop application-specific techniques, and address domain-driven questions.}

%{This seemingly odd behavior of node 9 highlights three important lessons. First, adopting a community-based approach to network analysis provides a vantage point to ask new research questions. 
%Second, one must be cautious when comparing the output of a community detection algorithm to known information on the network (frequently referred to as ``ground truth''), as the latter might include important additional information not captured by the network topology \citep{peel2016ground}.
%Finally, applied community detection should incorporate domain knowledge to choose appropriate methodologies, develop application-specific techniques, and address domain-driven questions.}

{With these lessons in mind, the rest of this Chapter is organized by the following case studies.}
In section~\ref{sec:dane}, we describe how communities {have been used to help predict which memes go viral on Twitter.}
In section~\ref{sec:peter}, we highlight political polarization in the U.S. {Congress,} {demonstrating} the use of communities to quantify polarization and identify node roles such as U.S. Senators that bridge the legislative space between political parties.
%in the political center that play in important role in bridging the party-centric groups legislative activity. 
%Detecting communities at multiple resolutions is a useful tool that can reveal new hidden functionality and provide domain experts with possible research directions. 
%
In section~\ref{sec:clara}, we present a study of the neuronal network of \emph{C. elegans} in which multiresolution communities uncover groups of neurons with similar biological function.
In section~\ref{sec:saray}, we {turn to a different neuroscience application} that uses communities
%are associated with people's tasks} 
to compare human brain networks under different tasks and rest states. 
Finally, in section~\ref{sec:natalie} we provide an example of how {communities can help} explain {the evolution of genes important to Malaria.}
%help us understand the underlying mechanisms of biological processes. 
 %
%Our aim in this review is to describe how community detection has aided in the analysis of network data in a variety of examples across disciplines. 
%
%%%
%\textcolor{blue}{NS: The sentences I colored in red seem a bit informal.idk though? my suggestion is the black text}%%%%%
%\textcolor{red}{In making our choices about what to include here, we are admittedly highly-biased (and self-absorbed) by our own personal perspectives about the development, impact, and future direction of community detection in networks. Our viewpoint is of course not comprehensive and we apologize for the many important contributions which we do not make space to address in the present work. But we are hopeful that newcomers (and perhaps also some who are not so new to the area) will find our discussion to be useful and/or thought provoking.}\todo{Dane agrees}
%
%While these examples span multiple disciplines, they by no means provide a complete discussion of community detection.
% of successful \drt{uses} of community detection. 
{We selected these case study examples to highlight}
%We sought to motivate 
the utility {of a} community-driven approach to network analysis, drawing from these creative applications {in which} the {modeling} assumptions {and algorithm choices elucidate} important aspects of the data. We hope that our discussion will be thought provoking for those previously unfamiliar with {this area} and inspire {further use of community detection for network analysis.}

%%%%%%%%%%%%%%%%%%%%%%%%%%%%%%%%%%%%%%
\section{Virality prediction of social memes}\label{sec:dane}

Community structure affects social contagions and epidemics through structural trapping, meaning that a meme or virus spreads readily within a community (or communities, if the contagion arises in clusters)
%\citep[or communities, if the contagion arises in clusters][]{taylor2015topological}
and tends to not spread (as quickly, if at all) from one community to another \citep{onnela2007structure,aral2009distinguishing,mcpherson2001birds,melnik2014dynamics,o2015mathematical,ugander2012structural,weng2013virality}. That is, the contagion exhibits ``{community concentration}'' in which it is localized (i.e.~concentrated) within one or more communities. In the context of epidemics, structural communities (which often reflect geographic constraints) can be represented by metapopulation models \citep{colizza2008epidemic,melnik2014dynamics} that partition the human population into subgroups (broadly defined).
Social contagions and epidemics share many mathematical and modeling similaritities \citep{dietz1967epidemics,goffman1964generalization}; however, their differences are also important. One crucial distinction is that social contagions are typically better modeled as {complex contagions} \citep{centola2007complex} in which a node's (i.e.~person's) adoption of the contagion requires \emph{social reinforcement}, e.g., as modeled by a threshold criteria \citep{granovetter1978threshold}.
Whereas a biological epidemic can be transmitted through a single exposure, a person can require a certain amount of ``contagion exposure'' (e.g., number or fraction of contacts who have already adopted it) before adopting a social contagion themselves. Although subtle, this discriminating feature of social contagions and epidemics can significantly impact spreading patterns on networks \citep{centola2010spread,centola2007complex,taylor2015topological,weng2013virality,Melnik2013,o2015mathematical}.
%That is, before adopting a contagion themselves, a person may require a certain amount of  ``contagion exposure''  (e.g., number or fraction of contacts who have already adopted it). Although subtle, this discriminating feature of social contagions and epidemics can significantly impact spreading patterns on networks \citep{centola2010spread,centola2007complex,taylor2015topological,weng2013virality}. 

\citet*{weng2013virality} study the spread of memes across the Twittersphere, concluding that homophily and social reinforcement collectively boost community concentration. Interestingly, they find this effect to differ for viral memes (those that spread vastly in the population) versus non-viral memes (those that do not reach high levels of popularity and are only shared by a small fraction of the population).
%%%appear in a large number of messages and are adopted by many people---versus non-viral (i.e., typical) memes.
%%%They identify three central contributions of their work: 
The three main findings of their work are:
(I)~communities allow us to estimate how much the spreading pattern of a meme deviates from that of infectious diseases; 
(II)~viral memes tend to spread more like epidemics than non-viral memes; and finally
%%%(III)~[they] predict the virality of memes based on early spreading patterns in terms of community structure. We now describe further each of these results.
(III)~the virality of memes can be predicted based on early spreading patterns in terms of community structure. We now describe further each of these results.

%%% EXTRACTED FROM THE METHODS SECTION: We collected a 10% sample of all public tweets from Mar 24 to Apr 25, 2012 using the Twitter streaming API (dev.twitter.com/docs/streaming-apis). Only tweets written inEnglish are extracted. The dataset comprises 121,807,378 tweets generated by 14,599,240 unique users, and containing at least one of 10,393,465 hashtags. We then constructed an undirected, unweighted network based on reciprocal following relationships between 595,460 randomly selected users, as bi-directional links reflect more stable and reliable social connections

%%%The authors study a network encoding following among \pjm{[Previous three words left me unsure]} Twitter users and providing evidence of structural trapping for memes, defined as unique hashtags, that spread through tweets and retweets. 
%%%The authors study a network where vertices represent Twitter users and edges account for reciprocal following relationships
The authors built an unweighted, undirected network from Twitter data, encoding reciprocal following relationships between users. This network provided evidence of structural trapping for memes, defined as unique hashtags, that spread through tweets and retweets. They identified communities using two community-detection methods: Infomap \citep{rosvall2008maps}, an information-theoretic algorithm; and link clustering \citep{ahn2010link}, which identifies overlapping communities by clustering edges. By analyzing the flow of information, they found that memes are much more likely to spread across intra-community edges 
%(i.e.~inside communities)
versus inter-community edges.
% (i.e.~between communities)
Given that a variety of factors (e.g., homophily, social reinforcement, and use history) can contribute to this phenomenon, it is important to recognize that this feature of community structure alone is able to differentiate how important different edges might be in fostering the spread of memes.

To demonstrate that the local phenomenon of preferential spreading across intra-community edges contributes to the mesoscopic phenomenon of community concentration, the authors developed an entropy-based measure to quantify the extent to which the spreading of memes concentrates into communities. They compared this measure for their data set to that of four null models for social contagions: random spreading, a simple epidemic, a social reinforcement model, and an epidemic with homophily. By drawing this comparison, the authors observed community concentration for non-viral epidemics to more-closely resemble complex contagions, whereas the spreading of viral memes %viral epidemics 
more-closely resembled simple epidemics. In particular, viral memes exhibited less structural trapping (similar to epidemics), whereas non-viral memes exhibited stronger structural  trapping (similar to complex contagions).

To further distinguish viral and non-viral memes, Weng et al. focused on the early stages of contagions and studied the average contagion exposure (i.e.~the number 
% or fraction \pjm{[Just checking: they do both?]} 
of social contacts who are already adopters) for each adopter of a contagion. The authors compared their Twitter data set to the same four null models and again observed viral memes to more-closely resemble simple epidemics; namely, less exposure is required for transmission of a viral meme.

Motivated by the observation that community concentration and contagion exposure are informative features to gauge the virality of a meme, they then implemented a classification algorithm using random forests to predict whether or not a meme will go viral. To map virality prediction as a classification problem, they partitioned the set of memes into two classes (viral versus non-viral) so that they can specify the fraction of memes that are non-viral (considering virality both in terms of the number of retweeters and the total number of retweets). To study the benefit of using community structure information to improve virality prediction, they compared the resulting classification precision and recall scores for three classifiers: random guessing, community-blind prediction and community-based prediction. They found that incorporating information about community structure can greatly improve the prediction accuracy for the virality of memes.

%%%%%%%%%%%%%%%%%%%%%%%%%%%%%%%%%%%%%%
\section{Congressional roll call}\label{sec:peter}
While the representation of Twitter following as a network is straightforward, direct connectivity is only one of many data types that can be represented by a network. Other common networks encode the similarity between, for example, people, text documents, or protein sequences. Here we consider communities found in network representations of roll-call voting similarity in the United States Congress, as constructed and studied by \cite*{waugh2009}. These networks connect two members in a selected Congress (that is, the two-year period starting in the early January following the biennial Congressional elections) according to the similarity in their voting patterns. \citet{waugh2009} defined edge weights equal to the fraction of bills that the two members voted the same way, yay or nay, among the total number of bills for which they were both present and voted (after removing nearly unanimous votes). This definition yielded weighted edges in a dense network; indeed, every member of Congress is connected in this definition to every other member in the same chamber with some positive weight unless they managed to never once vote the same way, while two members who always voted identically are connected with an edge of weight 1. Because the self-loops connecting each member of Congress to herself do not provide additional information, these were removed. This undirected roll-call-similarity network is a selected projection of the underlying bipartite (and signed) data that connects legislators with the bills that they voted on. This projection is useful for describing legislative activity because the community structures group together members of Congress who vote similarly, providing relatively accessible and intuitive examples of communities, independent of the political or policy content of the bills.
 %%%(at least for those readers who follow or are aware of some of the history of U.S. politics).

\begin{figure}[h]
\begin{center}
\includegraphics[width=0.45\textwidth]{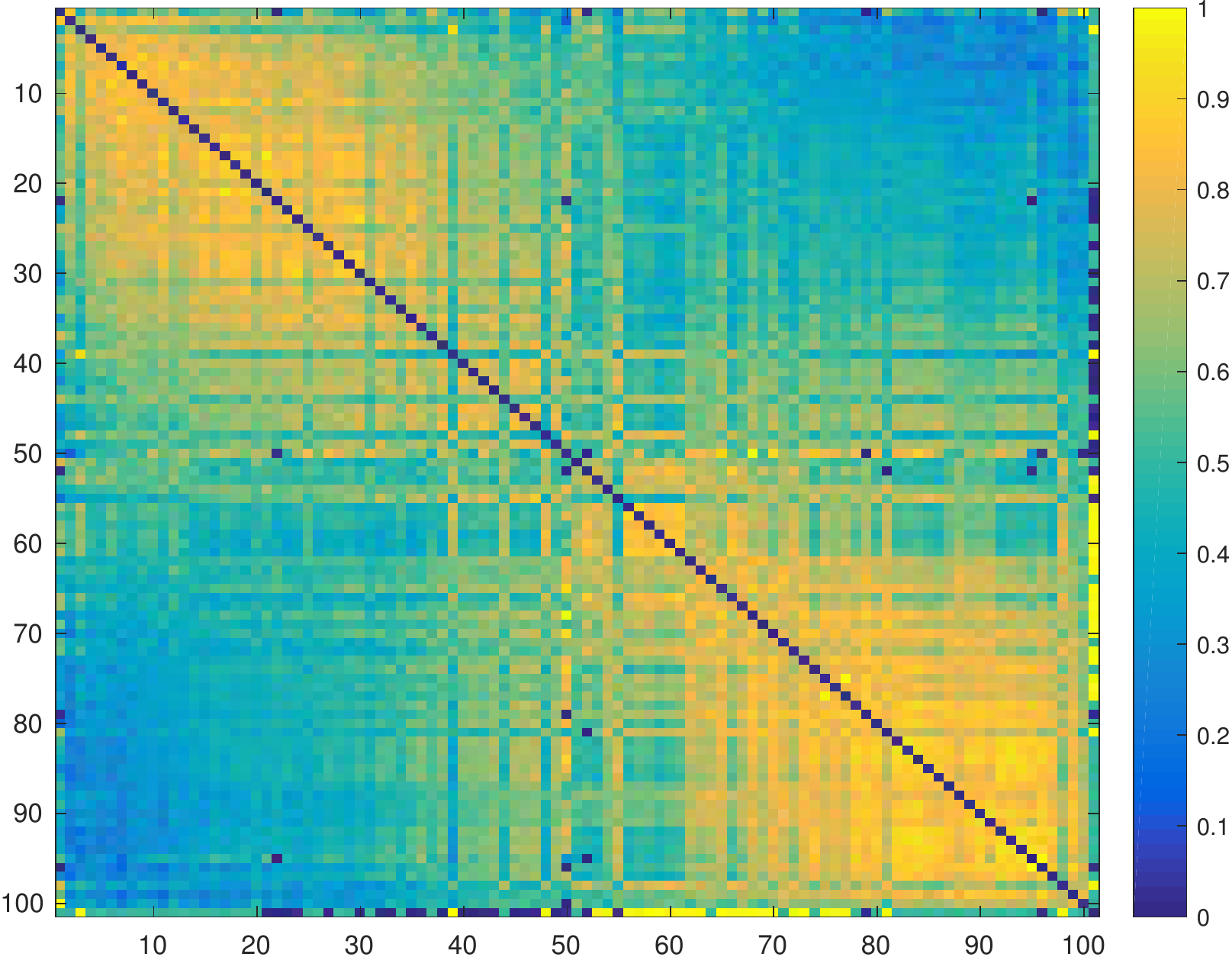}\qquad\includegraphics[width=0.45\textwidth]{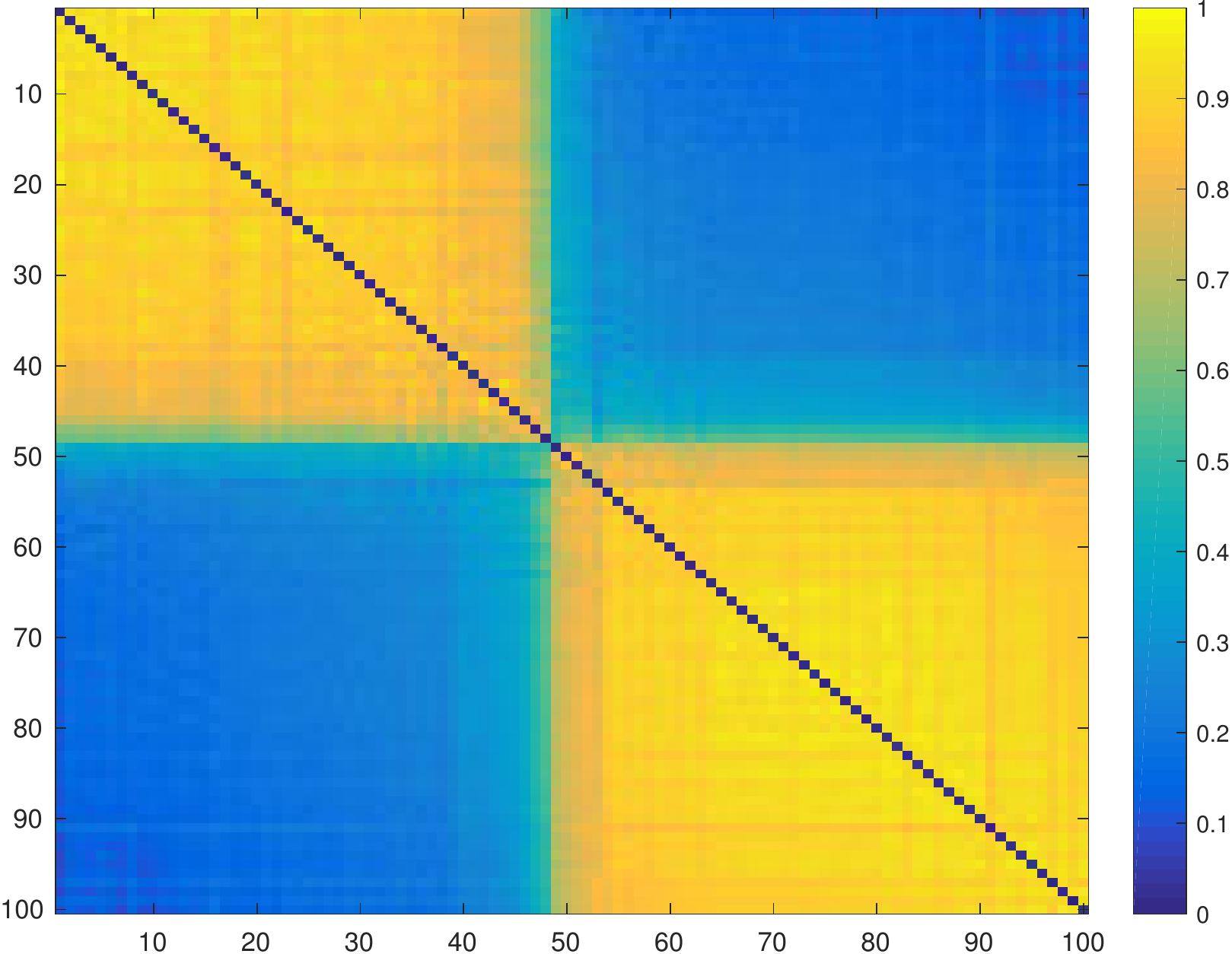}
\caption{Roll call similarity adjacency matrices in the U.S. Senate as defined by \cite{waugh2009} for the (left) 85th and (right) 108th Congresses, after reordering indices (Senators) with reorderMAT from the Brain Connectivity Toolbox \citep{Rubinov2010}. The 85th Congress, January 3, 1957 -- January 3, 1959, included the first federal civil rights legislation passed by Congress since Reconstruction \citep{CivilRightsAct1957}. The modularity of this weighted network (that is, the maximum modularity obtained across observed partitions) is 0.091. In contrast, the modularity of the 108th Senate,  January 3, 2003 -- January 3, 2005, is 0.273, one of the highest values in any Senate. For comparison, two equal-sized blocks with perfect agreement within and zero agreement between blocks yields a modularity of $1/2$ (up to a $1/N$ factor from removal of self loops).}
\label{fig:rollcall}
\end{center}
%\vspace{-.5 cm}
\end{figure}

\citet{waugh2009} studied community structure for these networks, providing a framework for thinking about the large-scale structure of Congressional legislative action in terms of political allegiances, whether or not those allegiances are well aligned with the nominally declared party memberships.
%As argued by \citet{waugh2009}, the modularity of the roll call similarity network (that is, the largest value of modularity found maximizing over partitions) provides a useful measure of legislative polarization. In particular,  they found a curious relationship between the modularity of a chamber (House or Senate) in a Congress and turnover of its majority party in the elections leading into the next Congress:
In particular, they considered modularity, which measures the difference between the total weight of within-community edges and that expected under a given null model (e.g. random network), to quantify legislative polarization. They found a curious relationship between the modularity of a chamber (House or Senate) in a Congress (that is, the largest value of modularity found maximizing over partitions) and turnover of its majority party in the elections leading into the next Congress:
while periods of very high or very low polarization (as measured by modularity) appeared to be relatively stable in terms of re-electing the majority party, they found that middle levels of polarization more frequently led to majority party turnover at the subsequent election (controlling for various other hypothesized factors). In so doing, \citet{waugh2009} not only used community detection as an exploratory tool for intuitively understanding large-scale structure, they additionally used modularity as a useful quantity describing an important global feature of these networks.

Additional intuition about these networks and their changes over time can be obtained from visualizations, as demonstrated by the force-directed layouts of \citet{andris2015} and the community-focused figure of \citet{moody2013}.
%\footnote{The \citet{moody2013} visualization is also available at \url{https://www.washingtonpost.com/news/the-fix/wp/2013/06/05/it-wasnt-always-this-bad-the-growth-of-political-polarization-in-1-chart/} .} 
Looking at the Senate roll call from 1975 to 2012, \citeauthor{moody2013} combined modularity for system-wide polarization of a Congress with groups of Senators in each Congress identified by a modified version of ``convergence of iterated correlations" \citep[CONCOR,][]{breiger1975,white1976}. A feature of this grouping is that by construction it leaves some Senators in the political center unaffiliated with the party-centric groups, allowing for easy visualization of the hollowing out of legislative activity in the political center over time, along with increasing polarization. This simultaneous use of modularity and CONCOR in the visualization demonstrates the value of using multiple methods for identifying communities.

Because of the temporal nature of the roll call networks, community-detection methods that explicitly utilize the identifications across time have also been usefully applied. Whereas the \citet{waugh2009} analysis and \citet{moody2013} visualization detected communities within each two-year Congress independently (and then identify common Senators from one Congress to the next in the visualization), a ``multilayer networks" framework can be used for studying networks that change dynamically over time, as well as a variety of other network generalizations \cite[see, e.g.,][]{kivela2014}. \citet{mucha2010community} used the properties of Laplacian dynamics to generalize the original definition of modularity to multilayer networks, using the Senate roll call as an instructive example for temporal networks, processing the data into the two-year single-Congress waves (called ``slices" then but now more commonly thought of as ``layers"). Naively, one could start by maximizing the modularity of each layer independently, but connecting those communities between layers then requires selection of a matching procedure that often leads to ambiguities. In contrast, the multilayer version directly allows for continuation of communities from one layer to the next and characterizing their flow across layers. In the simplest setting, the idea behind multilayer community detection introduces an interlayer coupling parameter, $\omega$, describing the weight of the identity arcs linking corresponding nodes across layers.  The multilayer modularity and the partitions found under fixed parameters then depend on $\omega$.
For $\omega=0$, the single-layer modularity of each network layer is optimized independently. %Thus, the resulting partition will simply correspond to the union of the independent partitions obtained from optimizing the standard modularity function.
As $\omega$ is increased, the coupling between layers encourages finding partitions that include greater spanning of communities across layers. 

The partition highlighted and visualized in \citet{mucha2010community} includes communities that span multiple Congresses, with most of the single-Congress layers containing only two communities. The handful of layers with more than two communities mark key transitions in the two-party system, often with one group fading in favor of another (whether or not they name themselves differently). While the start of the American Civil War in 1861 is particularly obvious in the data, these transitions also occur near other major political moments or, in some cases, near the boundaries of the recognized ``Party Systems" of the United States as studied in political science \cite[see][]{PoliticalParties}. Alternative partitions of the data corresponding to different interlayer coupling parameter values were visualized by \citet{muchaporter2010}, demonstrating how different features are highlighted by exploring the space of community detection parameters. 

We note that similar network constructions have been used to study voting in the Congresses in Peru \citep{lee2017} and Brazil \citep{levorato2016}, as well as the United Nations General Assembly \citep{macon2012}. Community detection has also been used to study committee assignments \citep{Porter_Mucha_Newman_Friend_2007,Porter_Mucha_Newman_Warmbrand_2005,Porter_Friend_Mucha_Newman_2006} and cosponsorship \citep{Zhang_Friend_Traud_Porter_Fowler_Mucha_2007} in the U.S. Congress, and multilayer modularity in the multiplex setting was used by \citet{cranmer2015} to measure the level of ``fractionalization" in international relations.

%\clearpage

%%%%%%%%%%%%%%%%%%%%%%%%%%%%%%%%%%%%%%
\section{Exploratory analysis of the C. elegans neural network}\label{sec:clara}

Many community-detection methods, including but not limited to many traditional modularity optimization algorithms, provide a user with a single partition of the network into communities along with the corresponding value of the objective function (e.g., modularity). The value of modularity itself can be valuable as in the example of the previous section and is frequently interpreted by users as an assessment of the meaningfulness of that partition, although caution is strongly recommended \citep[see][]{guimera2004,bassett2013}. There are two immediate problems with analyzing a network with a fixed resolution community-detection algorithm. First, some meaningful structures could remain undetected (e.g. small cliques lumped together into one community) under modularity optimization at a single resolution \citep{Reichardt_Bornholdt_2006} due to resolution limits of modularity \citep{fortunato2007resolution}, as well as detectability limits that apply to all polynomial-time community-detection methods \citep{nadakuditi2012}. Second, when the purpose of community detection is data exploration, studying a single resolution (or scale) of community structure might lead to the conclusion that there is only one good way to partition that data (which is often misleading). Instead, being able to access multiple scales of resolution of the data can be crucial for identifying and understanding interesting phenomena that otherwise would have been unexplored.

One example that illustrates the importance of multi-resolution community detection is a study of the neural network of the nematode \emph{Caenorhabditis elegans}. \emph{C.~elegans} is a free-living, transparent nematode that has become one of the most widely studied living organisms in biology. \emph{C.~elegans} was the first multicellular organism to have its whole genome sequenced and is currently still the only organism for which we have access to its whole connectome. %\pjm{Double-checking, is this still true?}. Clara: It seems so, there is progress in studying the connectome of the drosophila melanogaster but still they don't know the full system. I think we're safe.
The structural anatomy of \emph{C.~elegans} is approximately a cylinder of diameter 0.1~mm and length 1~mm. The structure of its neuronal wiring can be found in the \emph{Wormatlas} database~\citep{wormatlas}, consisting of 302 neurons, their locations, and the synapses between them as determined by serial section electron microscopy. The database also describes different functions in which each neuron is involved.

%\todo{Item for the group: eventually we have to pick a style about naming names or saying, e.g., ``the authors" at the start of the next paragraph, but we had better check with Jim and Ryan about the citation style of the Handbook, since it might make the choice for us if that style is to cite by author name}

\citet*{CElegans_sincro} and \citet*{cgranell_chaos} studied the structure of the nematode from a complex networks perspective, illustrating that community analysis can help discern the interplay between the topology and functionality of neural networks. The network abstraction describes the nervous system of \emph{C.~elegans} as a directed, weighted network, where nodes represent neuronal cell bodies and edges represent synapses. The resulting network was analyzed via modularity optimization
\citep[using the original formulation of][]{newmangirvan}, 
yielding a partition that divided the neurons into five communities corresponding mainly to locations on the worm's body. This result is not entirely surprising, as it indicates that synapses occur more often within identifiable spatially contiguous and determined regions as compared to a corresponding random-graph model (which is independent of spatial location).  
However, the authors were interested in analyzing the network at further resolution levels, in the
%hopes
hope that this would reveal new interesting features. To this end, \citet*{njp08} proposed an algorithm using a modified version of the original modularity formulation, incorporating a tuning parameter to detect communities across the whole mesoscale. This was done by adding a self-loop of equal weight $r$ to all nodes in the network, a modification that only affects the diagonal of the adjacency matrix and therefore keeps the network connectivity unchanged \citep[cf. the different resolution parameter approach introduced by][]{Reichardt_Bornholdt_2006}.
When the weight $r$ takes its minimum value, the maximum-modularity partition for this modified network is a single community including all nodes (the \emph{macroscale}). Conversely, when the weight of the self-loop is tuned to its maximum value, the corresponding partition separates each node into its own community (the \emph{microscale}). By tuning $r$ between these two extreme values, one can explore community structure at different resolutions. It is worth noting that as each modularity optimization is independent from the others, the obtained structure is not forced to follow a hierarchical structure.
%\citep[Other multiresolution algorithms, such as][have been developed to identify hierarchal levels of communities akin to hierarchal clustering]{Reichardt_Bornholdt_2006}.
%\citep[Alternatively, using the methodology of][communities across multiple resolutions can be obtained with different values of the resolution parameter.]{Reichardt_Bornholdt_2006}

To apply this algorithm to the \emph{C.~elegans} neural network, \citet{cgranell_chaos} discretized the self-loop weight range into 1000 logarithmically spaced intervals, spanning $r \in [0,r_{\mbox{\scriptsize max}}]$. By considering $r>0$, they tunably identified a greater number of communities (whose sizes decreased) with increasing $r$. The mesoscale is depicted in Fig.~\ref{fig:panel}(A), where we can observe multiple important resolution scales. The most persistent scale of community structure is highlighted by a circle in the figure, providing evidence that at this scale the communities are robustly detected.
To simultaneously extract information across scales, they built a frequency matrix (or ``consensus matrix") encoding the number of times that two neurons were placed in the same community for the different $r$ values. By thresholding these frequencies, they were able to unravel sub-structural scales corresponding to groups of neurons involved in different functionalities at different scales. Figure~\ref{fig:panel}(B) shows the frequency matrix thresholded at 0.6, a value chosen by fixing the sizes of the groups to be analyzed to ten neurons or less. The figure highlights the five large communities corresponding to optimizing the original modularity measure (i.e.~$r=0$), as well as the substructures within these five communities. In particular, the highlighted scales in Fig.~\ref{fig:panel}(A) contributed most to the frequency matrix.

\begin{figure}[t]
\begin{center}
\begin{tabular}[t]{p{5mm}cp{5mm}c}
   \begin{tabular}[b]{c}
    A) \\ \mbox{\rule{0pt}{137pt}}
    \end{tabular}
    &
    \mbox{\includegraphics[height=0.26\textheight]{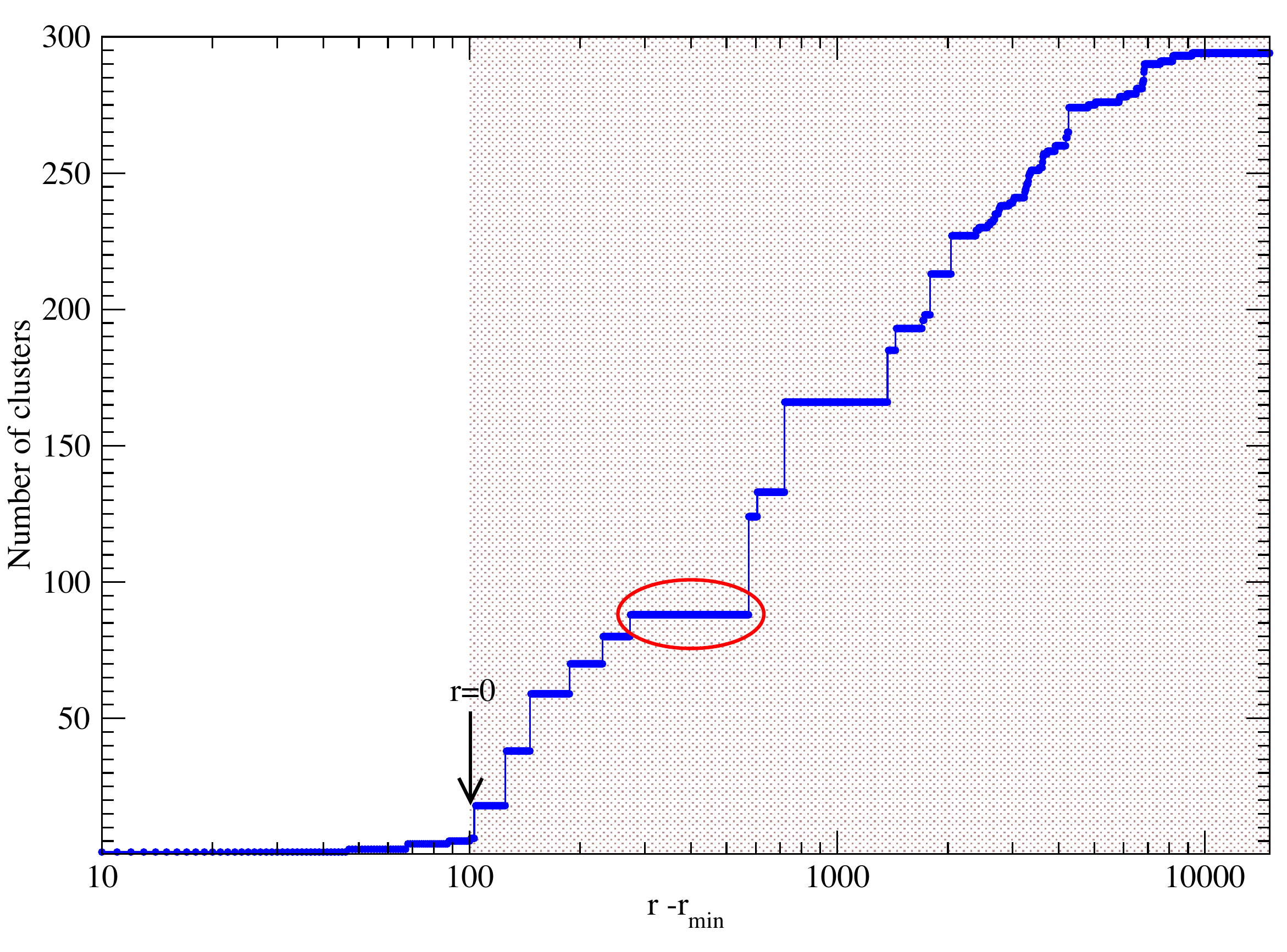}}
    &
   \begin{tabular}[b]{c}
    B) \\ \mbox{\rule{0pt}{137pt}}
    \end{tabular}
    &
    \mbox{\includegraphics[height=0.26\textheight]{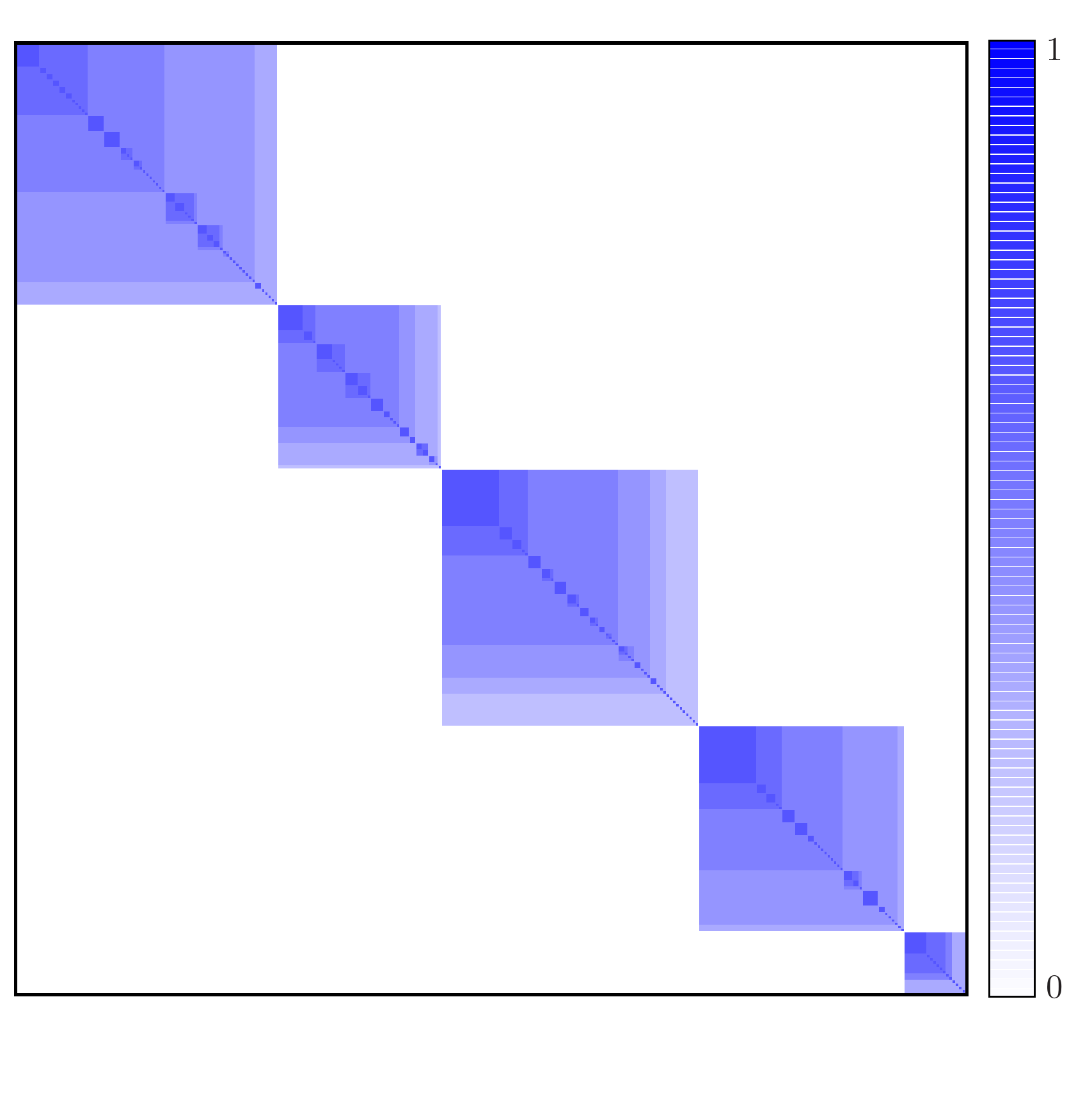}}
  \end{tabular}
\caption{\small Results of a multiresolution-community-detection algorithm for the \emph{C.~elegans} neural network. Panel A shows the number of detected communities for the modularity-optimizing partition at every value of the topological scale defined by the $\log (r-r_{\scriptsize\mbox{min}})$, where $r_{\scriptsize\mbox{min}}$ is the value of $r$ that maximizes the modified modularity measure for the macroscale-partition (i.e., the partition obtained with $r=0$).
%is the exact value of the self-loop that makes the macroscale optimal to modularity.
Panel B visualizes the frequency matrix of the mesoscales of the \emph{C.~elegans}, thresholded at a value of 0.6.}
\label{fig:panel}
\end{center}
\end{figure}

Trying to classify the functional role of neurons in \emph{C.~elegans} is extremely delicate because of their multifunctional aspects; that is, many neurons participate in different synaptic pathways resulting in different functionalities. However, with the previously obtained partition and the extensive description of each neuron in the \emph{Wormatlas} database, \citet{cgranell_chaos} proposed a tentative classification of some groups of neurons. The task involved assigning functions to groups of nodes that are persistently co-clustered across many scales of resolution. They identified nine groups of neurons that were both strongly persistent and small (specifically, they contained fewer than 10 neurons) and found these communities to be strongly associated to the following functional roles: (I) nose/head orientation movement; (II) head-withdrawal reflex, related to dorsal relaxation; (III) head-withdrawal reflex, related to ventral relaxation; (IV) olfactory and thermosensation reflex; (V) chemotaxis to lysine reflex; (VI) backward sinusoidal movement of the worm, related to touch stimulus; (VII) forward and backward autonomous sinusoidal movement of the worm; (VIII) relaxation state related to a sleep state; and (IX) a group containing neurons with functions that remain unknown. Their classification does not intend to be exact or final, but rather to provide biologists with useful information for future research.

As we have seen, the application of community-detection algorithms is a powerful approach to exploratory data analysis. Moreover, the use of a multiresolution approach identified structures beyond the expected grouping of neurons in different locations, and allowed discovery of groups of neurons that contribute to the same neurological function, providing a takeoff point for further research.

%\clearpage
%%%%%%%%%%%%%%%%%%%%%%%%%%%%%%%%%%%%%%
\section{Comparing network architectures of the human brain at different states}\label{sec:saray}
%\input{sections/brain.tex}
%Things to mention in the inro:
%modularity quality function
% meso-scale
% partition: the assignment of nodes to communities is referred to as a ``partition''

%Another type of brain connectivity data is functional connectivity, 
Another type of neural-connectivity data is functional brain connectivity, which describes the statistical patterns of dynamic interactions among neurons or brain regions \citep{Bullmore2009}. Unlike the ``structural network'' in the previous section (where the network represents the actual wiring between neurons), 
``functional networks'' can be measured with a variety of neuroimaging or electrophysiological recording methods and can be measured while the brain is in a resting state or under stimulus \citep{Sporns2013}. 
The structural and functional brain networks of various model organisms (such as \emph{C. elegans} mentioned in Section \ref{sec:clara}) and humans have been shown to organize into communities (usually called modules in this context) %(also called modules, especially in this context) 
which often correspond to specialized functional components \citep{Sporns2016}.
Such a modular organization has been suggested as evolutionarily advantageous for several reasons. For instance, it conserves the wiring cost involved in anatomically connecting neurons to constitute circuits or networks, since the connections inside communities are often shorter \citep{Bullmore2012}.
Moreover, changes in the modular organization of the human brain have been recently shown to associate with aging and clinical disorders \citep{Fornito2015}.

\begin{figure}[h!]
\begin{center}
\includegraphics[width=0.45\textwidth]{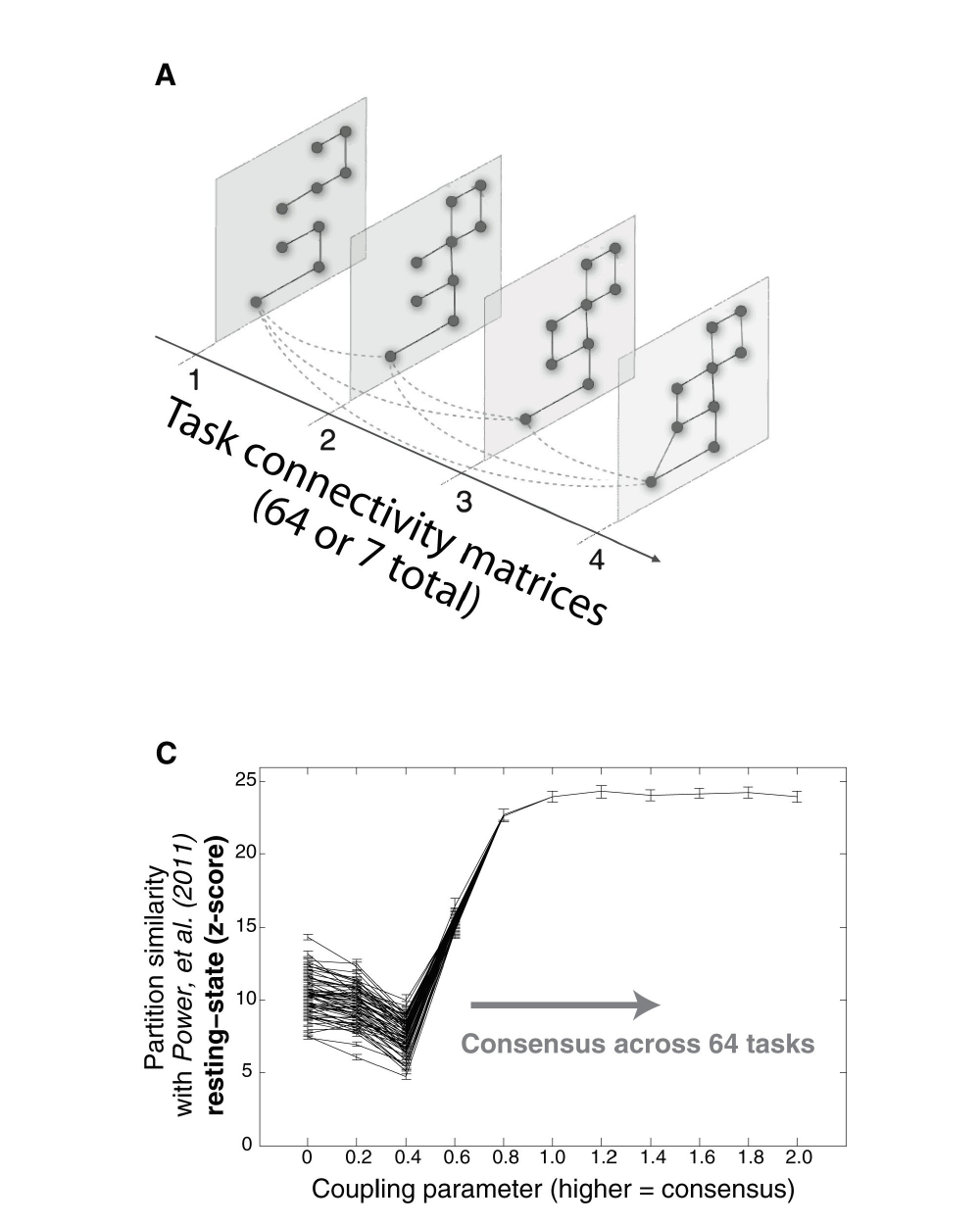}\hspace{0.2em}\includegraphics[width=0.45\textwidth]{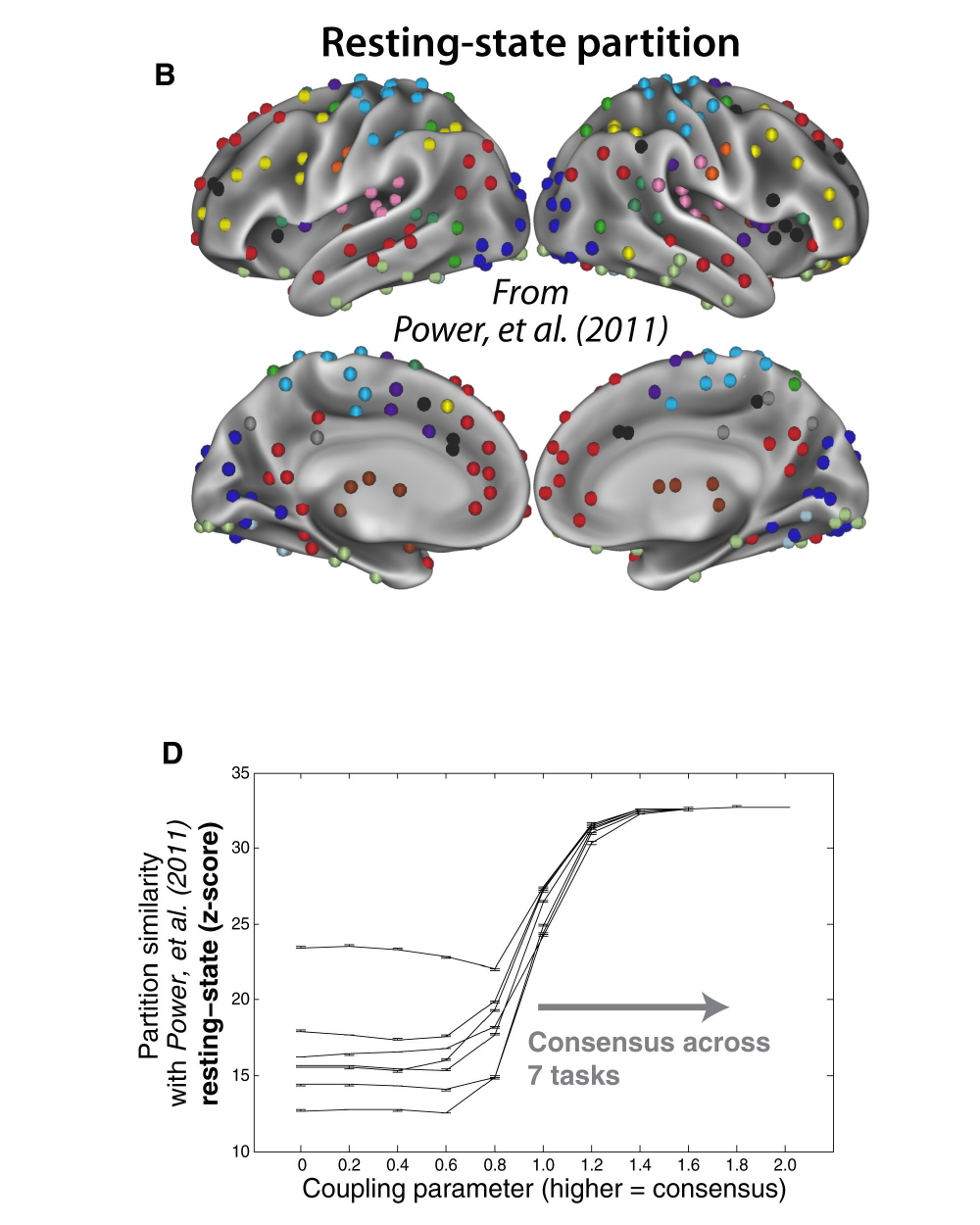}
\caption{Multilayer community detection applied to resting state and task-evoked functional brain networks \citep[Reprinted figure with permission from][\textcopyright 2014 Elsevier Inc]{Cole2014}. Each layer in the multilayer network (schematic shown in panel A) represents the functional connectivity between brain regions under different tasks. The layers are coupled by identity arcs of weight $\omega$ connecting each node (brain region) in a given layer to itself in all other layers (dashed lines). Panels C-D show the similarity (measured by the standardized Rand coefficient) of each task partition to the resting-state partition reported in \citet{Power2011} (shown in panel B) as a function of $\omega$. As $\omega$ increases, the task partitions converge to a consensus partition similar to the resting-state community partition.}
\label{fig:cole}
\end{center}
\end{figure}

However, community detection applied to a static single network can fail to capture more realistic situations where the data is temporal, originates from multiple sources or spans multiple spatial and/or temporal scales.
To address this shortcoming, some community detection techniques have been recently extended for \textit{multilayer networks} in which multiple networks form a multilayer stack as shown in Fig.~\ref{fig:cole}{A} (see also the discussion in Section \ref{sec:peter} on the use of multilayer networks in studying temporal Senate roll-call networks).
In general, these layers can represent different time windows in an experiment, different individuals, or different experimental conditions.% In this particular example, different layers account for snapshots taken from the same network at different times. 

%Applying community detection to multilayer representations of brain networks can be used to track changes in functional modules over time (e.g. during learning and task execution~\cite{Bassett2011,Bassett2015}), compare modules in networks collected across multiple subjects or different experimental conditions~\cite{Cole2014}, and find communities that persist across multiple temporal or spatial scales~\cite{Betzel2016}.
%\pjm{``Multilayer'' networks were mentioned very briefly in the roll call section. The discussion of this paragraph is better (unsurprisingly). We need to be sure to merge appropriately so that they work together and reference each other.} \saray{does this look better?}

\citet*{Cole2014} applied multilayer community detection to characterize the relationship between resting-state (i.e.\ subjects were asked to do nothing) and task-evoked (i.e.\ subjects were asked to perform a specific task such as pressing a button or answering a logic question) functional connectivity in the human brain. Subjects were asked to perform different kinds of tasks while fMRI was used to measure the temporal changes in brain activity across hundreds of brain regions. Then, for each task, they constructed a layer in a multilayer network using the Pearson correlations between the fMRI time series of all pairs of brain regions. The authors hypothesized that networks obtained from resting-state fMRI would reveal an intrinsic architecture that would also be present across a wide variety of task states (i.e.\ across networks obtained from fMRI measurements under different tasks), but also that some task-evoked connectivity changes unique to each task state would be evident. 

To estimate both intrinsic and evoked architectures simultaneously, \citet{Cole2014} used the multilayer generalization of modularity
% \citep[][also discussed above in Section \ref{sec:peter} on Congressional roll call networks]{mucha2010community}
\citep{mucha2010community} to uncover communities spanning across layers.
In this setting, the multilayer formulation across subjects connects every brain region in a given layer to itself in each of the other layers with an identity arc of edge weight $\omega$, called the \emph{coupling parameter}. 
Note that in the temporal setting, as described in Section \ref{sec:peter}, each layer represents a different time window and each node is coupled to its appearances in consecutive \textit{ordered} layers; in contrast, here the layers are \textit{categorical} with all-to-all interlayer (intertask) connections.
%%% CLARA: I'm commenting this cause it's already solved, right?  \pjm{(Contrast this with the case of temporal networks, described briefly in the roll call Section above, where every layer represents a different time window and each node is coupled to its appearances between consecutive layers.)} \saray{great idea!}
%\clara{Either we move the whole explanation of the multilayer modularity method to the Congressional Roll Call or we change section order}.

The authors used small values of the coupling parameter, $\omega$ to identify network communities elicited differentially across tasks, and large values of $\omega$ to identify consensus communities present across tasks.
For a given $\omega$, they applied multilayer community detection and compared the partition obtained for each task layer with the resting-state partition reported by \citet{Power2011} \cite[which used the \textit{Infomap} community-detection method by][]{rosvall2008maps}.
In particular, for every value of $\omega$, they performed $100$ random optimizations and chose the one that was most similar on average to the other $99$ optimizations as the representative partition.
This is one example of a consensus algorithm, which is used to find stable results from a set of partitions delivered by stochastic methods, as encountered with some of the computational heuristics for modularity optimization.

The similarity between task-specific communities and resting-state communities is reported in Fig.~\ref{fig:cole}C-D, shown as a function of weight $\omega$.
Similarity was quantified by the z-score of the Rand coefficient, which counts the fraction of node pairs identified the same way by both partitions \cite[either together in both or separate in both, see][]{Traud2011}.
To ensure the robustness of the results, two data sets were used.
One data set consisted of $64$ tasks (each performed by $15$ individuals) defined as distinct cognitive processes with minimal perceptual changes across tasks. 
The second data set involves seven tasks (each performed by $118$ individuals) that were chosen to elicit the involvement of all major cognitive domains and brain systems.
In both data sets, it was found that as $\omega$ increases, a single architecture emerged with high similarity to the resting-state network architecture.
While multilayer community detection indeed encourages a single consensus partition at high coupling parameters, there is no guarantee that this partition would look like the resting-state partition. 
In other words, the network architecture present across many task states is also present during rest, implying an intrinsic network architecture.

Upon further examination, the authors identified a set of small (but likely functionally important) task-evoked connectivities that differed from the rest-state connectivities.
To quantify these network changes, they calculated the percentage of connections that significantly (quantified by \emph{t}-tests) changed from the rest state, revealing a prominent pattern of decreased within-community connectivity and increased between-community connectivity during task performance, which suggests a partial breakdown of network communities during task performance so that activity can better flow between systems with diverse functions.
 
Providing a mesoscale perspective on the organization of brain networks, multilayer community detection employed at different coupling parameters can be useful for network comparison. Here, the authors compared connectivity patterns between brain regions (representing the functional dependencies between their fMRI time series) under different tasks and a rest state, revealing an intrinsic community structure that was present across brain states as well as small (but consistent) changes in the community structures that were common across tasks.

\section{A probabilistic network model for malaria parasite genes}\label{sec:natalie}

In addition to the analysis of neuroscience data, community detection can be useful for analyzing other biological data. %For example, communities identified in a similarity network between cells in a tumor correspond to phenotypically-distinct cell types \cite{CellSort}. Alternatively, communities among networks of genes or proteins with a strong response to experimental perturbation can assist in prioritizing further experiment s\cite{MS}. 
The nature of community detection applied to biological data is desirable for developing a mechanistic understanding of the underlying system. Here we highlight the work of \cite*{larremore2013network}, which used community detection to develop and computationally investigate a hypothesis about the nature of recombination in the sequences of the genes (called \emph{var} genes) encoding proteins in the human malaria parasite \emph{Plasmodium falciparum} genome. This work is novel and interesting because the authors used a network representation of their data, along with the communities found in this representation, to formulate and validate biological hypotheses. 

Rich genetic diversity in the  \emph{var} genes of the human malaria parasite has been shown to contribute to the complexity of the epidemiology of the infection and disease. The parasite can change which of the  \emph{var} genes are expressed at any given time on the infected red blood cell, which prevents the antibody from recognizing and resisting the new protein. One diversity-generating mechanism is recombination, which is the exchange and shuffling of genetic information during mitosis and meiosis \citep{var}. The ability to understand genetic diversity is complicated by inadequate tools to uncover the phylogeny, or genetic relationship between sequences resulting from recombination events, in a scalable and statistically rigorous way. The typical analyses for evolutionary data assume a tree-like relationship between events, which is unrealistic for recombination data. To address this challenge, \citet{larremore2013network} use a novel approach: they cast their problem in terms of a collection of networks. Then, they apply community detection to each of the networks and use the properties of the communities to generate hypotheses of the mechanisms behind the recombination process.
%makes 3 novel contributions; 1) the authors cast their problem in terms of a collection of networks 2) Apply community detection to each of the networks 3) Use properties of the communities  identified within each network and the similarities of node-to-community assignments between networks to generate hypotheses of the mechanism behind recombination. 

More specifically, to investigate the heterogeneity and the corresponding possible patterns in recombination events across a set of 307 sequences from the \emph{var} gene, the authors restricted their analyses to 9 particular ``highly variable regions'' (HVR) within each of the 307 sequences. Then for each HVR, they constructed a network, where the nodes represented the 307 sequences and an edge was placed between a pair of nodes if they had evidence of a recombinant relationship, based on a notion of sequence similarity within the particular HVR. Communities were then identified in each of the 9 networks using a degree-corrected stochastic block model (SBM) approach \citep{degreeCorrect}. In the SBM, the probability of an edge existing between a pair of nodes depends on their community assignments and hence nodes within a community are connected to each other and to other communities in a characteristic way. For a network with $N$ nodes and $K$ communities, the SBM is parametrized by an $N$-length array ${\bf z}$, where $z_{i}$ gives the community assignment for node $i$, and a $K \times K$ matrix, ${\boldsymbol \theta}$, where $\theta_{ij}$ (together with the node degrees) specifies the probability of an edge existing between nodes in communities $i$ and $j$. In the process of fitting the SBM, one learns the parameters ${\boldsymbol \theta}$ and ${\bf z}$ that are most likely to describe the data, and hence these parameters can then be used to sample networks from the model. In this analysis, sampling from the model was useful because it allowed the authors to create synthetic networks to computationally validate their hypotheses about the constraints influencing recombination.

%The community detection results for a particular region called HVR6 are shown in figure \ref{MalariaComm}. Panel A shows the HVR6 network, where nodes are colored according to their inferred community assignment. Panel B gives the community-colored adjacency matrix, with the rows and columns sorted by the inferred community assignments. \clara{Natalie: you are repeating the same info in the text and in the caption. You can replace one of the duplicates with information about what do we learn from the visualization. For example, why are communities blue and green more connected among them while the red one has almost no connection to the other two. Or, what do these three communities actually mean?}

\begin{figure}
\begin{center}
\label{fig:MalariaComm}
\includegraphics[scale=.5]{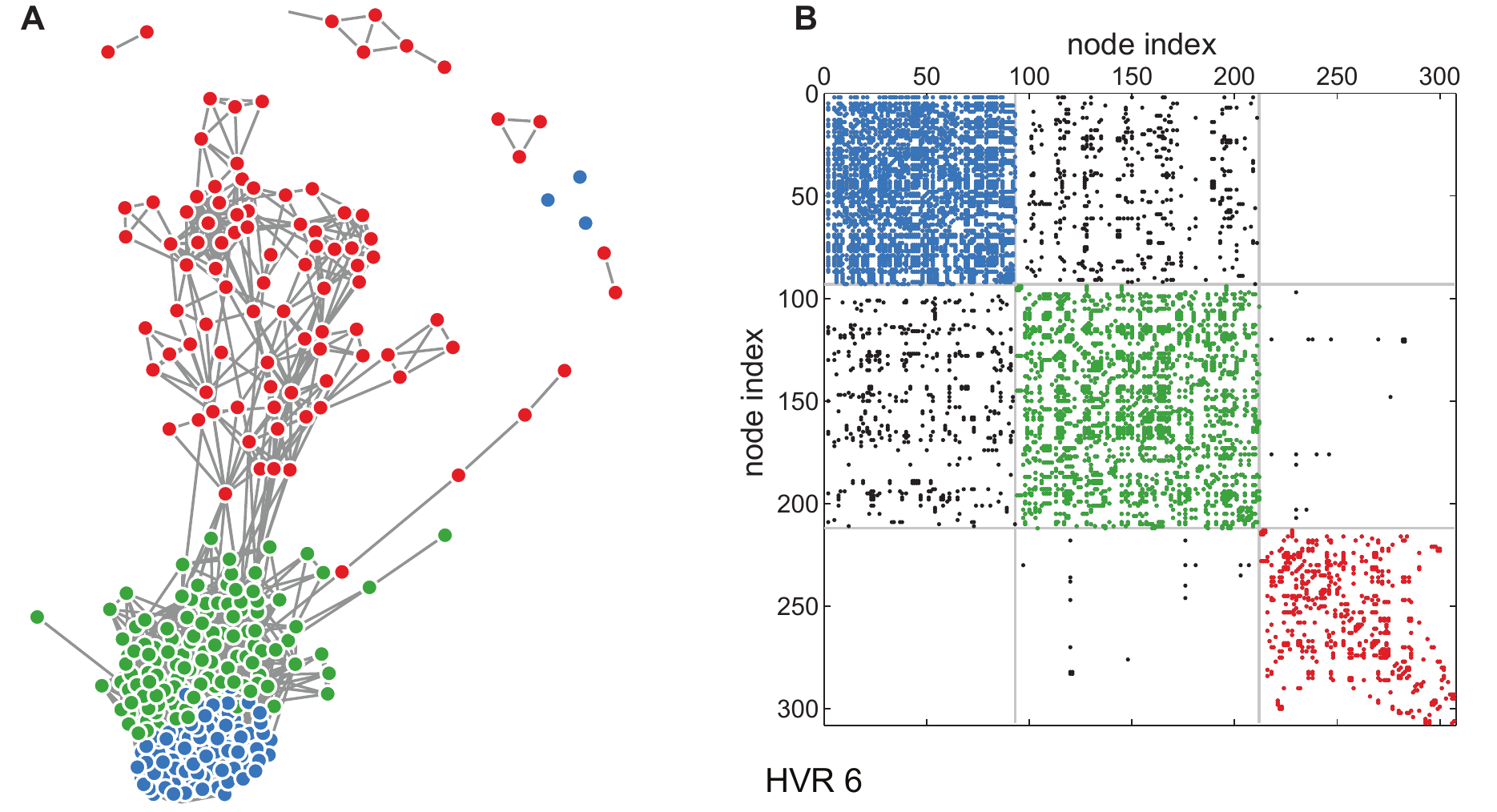}
\caption{Visualization of the community assignments inferred through the stochastic block model for HVR6 \citep[Reprinted figure with permission from][]{larremore2013network}. Panel A shows the HVR6 network, with the nodes colored by the inferred community assignment. Panel B gives the community-colored adjacency matrix, with the rows and columns sorted by the inferred community assignments.}
\end{center}
\end{figure}

%The assumption of the SBM is that the probability of an edge existing between a pair of nodes depends only on their community assignment and hence nodes with a community are connected to each other and to other communities in a characteristic way. Moreover, for a network with $N$ nodes, and $K$ communities, the SBM is parametrized by an $N$-length vector ${\bf z}$, where $z_{i}$ gives the community assignment for node $i$, and a $K \times K$ matrix, ${\boldsymbol \theta}$, where $\theta_{ij}$ gives the probability of an edge existing between nodes in communities $i$ and $j$. The SBM approach is useful here because in the process of fitting the SBM, one learns the parameters ${\boldsymbol \theta}$ and ${\bf z}$ that are most likely to describe the data, and hence these parameters can then be used to sample networks from the model. In this analysis, sampling from the model was useful because it allowed the authors to create synthetic networks to computationally validate their hypotheses about the constraints influencing recombination.

After identifying communities within each HVR network, as shown in Fig.~\ref{fig:MalariaComm}, the authors used two summary statistics to formulate their biological hypothesis. First, the variation of information \citep{VI} was used to compare the community assignments of nodes (i.e.\ each of the 307 sequences) across the 9 HVR networks. They observed that each network had a prominent community structure (i.e.\ far from random) and that the community assignments between networks were quite distinct. These observations motivated the hypothesis that recombination events occur in constrained ways, leading to a strong community structure, and that one should analyze HVR networks individually instead of building a consensus network that aggregates the HVR networks.  Next, they used \emph{assortativity} \cite[see, e.g.][]{assort} to overlay the network structure with various known biological features of the sequences, such as  \emph{var} gene length. Specifically, assortativity quantifies the tendency of nodes of the same type (e.g. same gene length) to be connected in the network. They observed that three HVR networks had community structure correlating strongly with two biological features (i.e. nodes of the same biological label tend to group together), while three other HVR networks with highly heterogenous community structure were unaligned with any of the known biology. These observations allowed for the formulation of the hypothesis that the HVRs that are unrelated to each other also promote recombination under unrelated constraints and are responsible for fostering genetic diversity to avoid immune evasion. 

Given the ability to find communities within each HVR network and the lack of similarity in community structure between HVR networks, \cite{larremore2013network} were able to formulate and test hypotheses for the diversity-generating mechanisms of \emph{var} genes, and this would have been difficult using standard phylogenetic approaches or without adopting a community-based perspective. The application of the stochastic block model to this task provided a statistically grounded approach for testing the plausibility of the model.

\section*{Concluding comments}
Through five representative case studies from diverse application domains, we have demonstrated the utility of community detection in data-analysis tasks such as prediction (see Section \ref{sec:dane}), node role classification and temporal evolution (see Section \ref{sec:peter}), multi-scale functional analysis (see Section \ref{sec:clara}), network comparison (see Section \ref{sec:saray}), and data representation for probabilistic model construction (see Section \ref{sec:natalie}). Our goal here was to provide the reader with an application-driven perspective on the various uses of community detection while highlighting application-specific goals and motivations for identifying communities in networks. We have by no means covered even a small fraction of the activity in community detection with the above examples, and many others could have been used [see, for example, recent applications in Hi-C data analysis \citep{Cabreros2016}, network security \citep{Ding2012} and understanding of animal societies \citep{Rubenstein2015}]. We hope that our presentation encourages readers to think about how community detection might be useful in their own work.
%\emph{This is a placeholder for now, seeking a good summary paragraph or two that captures the essences of the above examples and/or points the direction towards further developments in community detection.}

\section*{Acknowledgements}
%\todo{When we have a solid first draft, we want to share with some of our colleagues if we have time to see what feedback we get from them, and we can thank them here then.} 
The perspective on community detection presented in this work has been heavily influenced by the authors' research activities with many collaborators --- far too many to properly list here --- across a variety of supported projects and training grants from different agencies, most recently including the James S. McDonnell Foundation (grant \#s 220020315 and 220020457), the National Institutes of Health (Award Numbers R01HD075712, R56DK111930 and T32CA201159), and the National Science Foundation (ECCS-1610762). The present content is solely the responsibility of the authors and does not necessarily represent the official views of any funding agencies.

\small
%\bibliographystyle{abbrvnat}
%\bibliography{handbookrefs}

\begin{thebibliography}{106}
\providecommand{\natexlab}[1]{#1}
\providecommand{\url}[1]{\texttt{#1}}
\expandafter\ifx\csname urlstyle\endcsname\relax
  \providecommand{\doi}[1]{doi: #1}\else
  \providecommand{\doi}{doi: \begingroup \urlstyle{rm}\Url}\fi

\bibitem[Abbe(2017)]{Abbe2017}
E.~Abbe.
\newblock Community detection and stochastic block models: recent developments.
\newblock \emph{arXiv:1703.10146}, 2017.

\bibitem[Ahn et~al.(2010)Ahn, Bagrow, and Lehmann]{ahn2010link}
Y.-Y. Ahn, J.~P. Bagrow, and S.~Lehmann.
\newblock Link communities reveal multiscale complexity in networks.
\newblock \emph{Nature}, 466\penalty0 (7307):\penalty0 761--764, 2010.

\bibitem[Altun et~al.(2002)Altun, Herndon, Wolkow, Crocker, Lints, and
  Hall]{wormatlas}
Z.~Altun, L.~Herndon, C.~Wolkow, C.~Crocker, R.~Lints, and D.~Hall.
\newblock {Wormatlas. A database featuring behavioral and structural anatomy of
  Caenohabditis Elegans.}
\newblock \url{http://www.wormatlas.org/}, 2002.

\bibitem[Andris et~al.(2015)Andris, Lee, Hamilton, Martino, Gunning, and
  Selden]{andris2015}
C.~Andris, D.~Lee, M.~J. Hamilton, M.~Martino, C.~E. Gunning, and J.~A. Selden.
\newblock The rise of partisanship and super-cooperators in the u.s. house of
  representatives.
\newblock \emph{PLOS ONE}, 10\penalty0 (4):\penalty0 e0123507, 2015.

\bibitem[Aral et~al.(2009)Aral, Muchnik, and
  Sundararajan]{aral2009distinguishing}
S.~Aral, L.~Muchnik, and A.~Sundararajan.
\newblock Distinguishing influence-based contagion from homophily-driven
  diffusion in dynamic networks.
\newblock \emph{Proceedings of the National Academy of Sciences}, 106\penalty0
  (51):\penalty0 21544--21549, 2009.

\bibitem[Arenas et~al.(2006)Arenas, D{\'\i}az-Guilera, and
  P{\'e}rez-Vicente]{arenas2006synchronization}
A.~Arenas, A.~D{\'\i}az-Guilera, and C.~J. P{\'e}rez-Vicente.
\newblock Synchronization reveals topological scales in complex networks.
\newblock \emph{Physical Review Letters}, 96\penalty0 (11):\penalty0 114102,
  2006.

\bibitem[Arenas et~al.(2008{\natexlab{a}})Arenas, Fern{\'a}ndez, and
  G{\'o}mez]{CElegans_sincro}
A.~Arenas, A.~Fern{\'a}ndez, and S.~G{\'o}mez.
\newblock \emph{A Complex Network Approach to the Determination of Functional
  Groups in the Neural System of C. Elegans}, pages 9--18.
\newblock Springer Berlin Heidelberg, Berlin, Heidelberg, 2008{\natexlab{a}}.
\newblock ISBN 978-3-540-92191-2.

\bibitem[Arenas et~al.(2008{\natexlab{b}})Arenas, Fern\'andez, and
  G\'omez]{njp08}
A.~Arenas, A.~Fern\'andez, and S.~G\'omez.
\newblock Analysis of the structure of complex networks at different resolution
  levels.
\newblock \emph{New Journal of Physics}, 10\penalty0 (053039),
  2008{\natexlab{b}}.

\bibitem[Ball et~al.(2011)Ball, Karrer, and Newman]{Ball2011}
B.~Ball, B.~Karrer, and M.~E.~J. Newman.
\newblock Efficient and principled method for detecting communities in
  networks.
\newblock \emph{Physical Review E}, 84\penalty0 (3):\penalty0 036103, 2011.

\bibitem[Barnes(1982)]{Barnes1982}
E.~R. Barnes.
\newblock An algorithm for partitioning the nodes of a graph.
\newblock \emph{SIAM Journal on Algebraic Discrete Methods}, 3\penalty0
  (4):\penalty0 541--550, 1982.

\bibitem[Barry et~al.(2007)Barry, Leliwa-Sytek, Tavul, Imrie, Migot-Nabias,
  Brown, McVean, and Day]{var}
A.~E. Barry, A.~Leliwa-Sytek, L.~Tavul, H.~Imrie, F.~Migot-Nabias, S.~M. Brown,
  G.~A. McVean, and K.~P. Day.
\newblock Population genomics of the immune evasion (var) genes of plasmodium
  falciparum.
\newblock \emph{PLOS Pathogens}, 3\penalty0 (3):\penalty0 e34, 2007.

\bibitem[Bassett et~al.(2013)Bassett, Porter, Wymbs, Grafton, Carlson, and
  Mucha]{bassett2013}
D.~S. Bassett, M.~A. Porter, N.~F. Wymbs, S.~T. Grafton, J.~M. Carlson, and
  P.~J. Mucha.
\newblock Robust detection of dynamic community structure in networks.
\newblock \emph{Chaos}, 23\penalty0 (1):\penalty0 013142, 2013.

\bibitem[Binkiewicz et~al.(2014)Binkiewicz, Vogelstein, and
  Rohe]{binkiewicz2014covariate}
N.~Binkiewicz, J.~T. Vogelstein, and K.~Rohe.
\newblock Covariate assisted spectral clustering.
\newblock \emph{arXiv:1411.2158}, 2014.

\bibitem[Blondel et~al.(2008)Blondel, Guillaume, Lambiotte, and
  Lefebvre]{Blondel2008}
V.~D. Blondel, J.-L. Guillaume, R.~Lambiotte, and E.~Lefebvre.
\newblock Fast unfolding of communities in large networks.
\newblock \emph{Journal of Statistical Mechanics: Theory and Experiment},
  2008\penalty0 (10):\penalty0 P10008, 2008.

\bibitem[Bothorel et~al.(2015)Bothorel, Cruz, Magnani, and
  Micenkova]{Bothorel2015}
C.~Bothorel, J.~D. Cruz, M.~Magnani, and B.~Micenkova.
\newblock Clustering attributed graphs: models, measures and methods.
\newblock \emph{Network Science}, 3\penalty0 (03):\penalty0 408--444, 2015.

\bibitem[Breiger et~al.(1975)Breiger, Boorman, and Arabie]{breiger1975}
R.~L. Breiger, S.~A. Boorman, and P.~Arabie.
\newblock An algorithm for clustering relational data with applications to
  social network analysis and comparison with multidimensional scaling.
\newblock \emph{Journal of Mathematical Psychology}, 12\penalty0 (3):\penalty0
  328--383, 1975.

\bibitem[Bullmore and Sporns(2009)]{Bullmore2009}
E.~Bullmore and O.~Sporns.
\newblock Complex brain networks: graph theoretical analysis of structural and
  functional systems.
\newblock \emph{Nature Reviews Neuroscience}, 10\penalty0 (3):\penalty0
  186--198, 2009.

\bibitem[Bullmore and Sporns(2012)]{Bullmore2012}
E.~Bullmore and O.~Sporns.
\newblock The economy of brain network organization.
\newblock \emph{Nature Reviews Neuroscience}, 13\penalty0 (5):\penalty0
  336--349, 2012.

\bibitem[Cabreros et~al.(2016)Cabreros, Abbe, and Tsirigos]{Cabreros2016}
I.~Cabreros, E.~Abbe, and A.~Tsirigos.
\newblock Detecting community structures in hi-c genomic data.
\newblock In \emph{2016 Annual Conference on Information Science and Systems
  (CISS)}, pages 584--589, March 2016.
\newblock \doi{10.1109/CISS.2016.7460568}.

\bibitem[Centola(2010)]{centola2010spread}
D.~Centola.
\newblock The spread of behavior in an online social network experiment.
\newblock \emph{Science}, 329\penalty0 (5996):\penalty0 1194--1197, 2010.

\bibitem[Centola and Macy(2007)]{centola2007complex}
D.~Centola and M.~Macy.
\newblock Complex contagions and the weakness of long ties.
\newblock \emph{American Journal of Sociology}, 113\penalty0 (3):\penalty0
  702--734, 2007.

\bibitem[Cole et~al.(2014)Cole, Bassett, Power, Braver, and Petersen]{Cole2014}
M.~W. Cole, D.~S. Bassett, J.~D. Power, T.~S. Braver, and S.~E. Petersen.
\newblock Intrinsic and task-evoked network architectures of the human brain.
\newblock \emph{Neuron}, 83\penalty0 (1):\penalty0 238--251, 2014.

\bibitem[Coleman(1964)]{Coleman1964}
J.~S. Coleman.
\newblock \emph{Introduction to Mathematical Sociology}.
\newblock Collier-Macmillan, London, 1964.

\bibitem[Colizza and Vespignani(2008)]{colizza2008epidemic}
V.~Colizza and A.~Vespignani.
\newblock Epidemic modeling in metapopulation systems with heterogeneous
  coupling pattern: Theory and simulations.
\newblock \emph{Journal of Theoretical Biology}, 251\penalty0 (3):\penalty0
  450--467, 2008.

\bibitem[Cranmer et~al.(2015)Cranmer, Menninga, and Mucha]{cranmer2015}
S.~J. Cranmer, E.~J. Menninga, and P.~J. Mucha.
\newblock Kantian fractionalization predicts the conflict propensity of the
  international system.
\newblock \emph{Proceedings of the National Academy of Sciences}, 112\penalty0
  (38):\penalty0 11812--11816, 2015.

\bibitem[Delvenne et~al.(2010)Delvenne, Yaliraki, and
  Barahona]{delvenne2010stability}
J.-C. Delvenne, S.~N. Yaliraki, and M.~Barahona.
\newblock Stability of graph communities across time scales.
\newblock \emph{Proceedings of the National Academy of Sciences}, 107\penalty0
  (29):\penalty0 12755--12760, 2010.

\bibitem[Dietz(1967)]{dietz1967epidemics}
K.~Dietz.
\newblock Epidemics and rumours: A survey.
\newblock \emph{Journal of the Royal Statistical Society. Series A (General)},
  pages 505--528, 1967.

\bibitem[Ding et~al.(2012)Ding, Katenka, Barford, Kolaczyk, and
  Crovella]{Ding2012}
Q.~Ding, N.~Katenka, P.~Barford, E.~Kolaczyk, and M.~Crovella.
\newblock Intrusion as (anti)social communication: Characterization and
  detection.
\newblock In \emph{Proceedings of the 18th ACM SIGKDD International Conference
  on Knowledge Discovery and Data Mining}, KDD '12, pages 886--894, New York,
  NY, USA, 2012. ACM.
\newblock ISBN 978-1-4503-1462-6.
\newblock \doi{10.1145/2339530.2339670}.
\newblock URL \url{http://doi.acm.org/10.1145/2339530.2339670}.

\bibitem[Fiedler(1973)]{Fiedler1973}
M.~Fiedler.
\newblock Algebraic connectivity of graphs.
\newblock \emph{Czechoslovak Mathematical Journal}, 23:\penalty0 289--305,
  1973.

\bibitem[Fienberg and Wasserman(1981)]{Fienberg1981}
S.~E. Fienberg and S.~S. Wasserman.
\newblock Categorical data analysis of single sociometric relations.
\newblock \emph{Sociological Methodology}, 12:\penalty0 156--192, 1981.

\bibitem[Flake et~al.(2000)Flake, Lawrence, and Giles]{Flake2000}
G.~W. Flake, S.~Lawrence, and C.~L. Giles.
\newblock Efficient identification of web communities.
\newblock In \emph{Proceedings of the Sixth ACM SIGKDD International Conference
  on Knowledge Discovery and Data Mining}, KDD '00, pages 150--160, New York,
  NY, USA, 2000. ACM.
\newblock ISBN 1-58113-233-6.

\bibitem[Fornito et~al.(2015)Fornito, Zalesky, and Breakspear]{Fornito2015}
A.~Fornito, A.~Zalesky, and M.~Breakspear.
\newblock The connectomics of brain disorders.
\newblock \emph{Nature Reviews Neuroscience}, 16\penalty0 (3):\penalty0
  159--172, 2015.

\bibitem[Fortunato(2010)]{fortunato2010}
S.~Fortunato.
\newblock Community detection in graphs.
\newblock \emph{Physics Reports}, 486\penalty0 (3):\penalty0 75--174, 2010.

\bibitem[Fortunato and Barth{\'e}lemy(2007)]{fortunato2007resolution}
S.~Fortunato and M.~Barth{\'e}lemy.
\newblock Resolution limit in community detection.
\newblock \emph{Proceedings of the National Academy of Sciences}, 104\penalty0
  (1):\penalty0 36--41, 2007.

\bibitem[Fortunato and Hric(2016)]{fortunato2016}
S.~Fortunato and D.~Hric.
\newblock Community detection in networks: A user guide.
\newblock \emph{Physics Reports}, 659:\penalty0 1 -- 44, 2016.

\bibitem[Freeman(2004)]{Freeman2004}
L.~C. Freeman.
\newblock \emph{The Development of Social Network Analysis: A Study in the
  Sociology of Science}.
\newblock Empirical Press, Vancouver, Canada, 2004.

\bibitem[Galstyan and Cohen(2007)]{galstyan2007cascading}
A.~Galstyan and P.~Cohen.
\newblock Cascading dynamics in modular networks.
\newblock \emph{Physical Review E}, 75\penalty0 (3):\penalty0 036109, 2007.

\bibitem[Girvan and Newman(2002)]{Girvan11062002}
M.~Girvan and M.~E.~J. Newman.
\newblock Community structure in social and biological networks.
\newblock \emph{Proceedings of the National Academy of Sciences}, 99\penalty0
  (12):\penalty0 7821--7826, 2002.

\bibitem[Gleeson(2008)]{gleeson2008cascades}
J.~P. Gleeson.
\newblock Cascades on correlated and modular random networks.
\newblock \emph{Physical Review E}, 77\penalty0 (4):\penalty0 046117, 2008.

\bibitem[Goffman and Newill(1964)]{goffman1964generalization}
W.~Goffman and V.~Newill.
\newblock Generalization of epidemic theory.
\newblock \emph{Nature}, 204\penalty0 (4955):\penalty0 225--228, 1964.

\bibitem[Granell et~al.(2011)Granell, G\'omez, and Arenas]{cgranell_chaos}
C.~Granell, S.~G\'omez, and A.~Arenas.
\newblock Mesoscopic analysis of networks: Applications to exploratory analysis
  and data clustering.
\newblock \emph{Chaos}, 21\penalty0 (1):\penalty0 016102, 2011.

\bibitem[Granovetter(1978)]{granovetter1978threshold}
M.~Granovetter.
\newblock Threshold models of collective behavior.
\newblock \emph{American Journal of Sociology}, pages 1420--1443, 1978.

\bibitem[Granovetter(1973)]{Granovetter1973}
M.~S. Granovetter.
\newblock The strength of weak ties.
\newblock \emph{American Journal of Sociology}, 78\penalty0 (6):\penalty0
  1360--1380, 1973.

\bibitem[Guimer\`{a} et~al.(2004)Guimer\`{a}, Sales-Pardo, and
  Amaral]{guimera2004}
R.~Guimer\`{a}, M.~Sales-Pardo, and L.~Amaral.
\newblock Modularity from fluctuations in random graphs and complex networks.
\newblock \emph{Physical Review E}, 70\penalty0 (2), 2004.

\bibitem[Hastie et~al.(2001)Hastie, Tibshirani, and Friedman]{Hastie2001}
T.~Hastie, R.~Tibshirani, and J.~Friedman.
\newblock \emph{The Elements of Statistical Learning}.
\newblock Springer, Berlin, Germany, 2001.

\bibitem[Hastings(2006)]{Hastings2006}
M.~B. Hastings.
\newblock Community detection as an inference problem.
\newblock \emph{Physical Review E}, 74\penalty0 (3):\penalty0 035102, 2006.

\bibitem[Holland et~al.(1983)Holland, Laskey, and Leinhardt]{Holland1983}
P.~W. Holland, K.~B. Laskey, and S.~Leinhardt.
\newblock Stochastic blockmodels: First steps.
\newblock \emph{Social Networks}, 5\penalty0 (2):\penalty0 109--137, 1983.

\bibitem[Jeub et~al.(2015)Jeub, Balachandran, Porter, Mucha, and
  Mahoney]{Jeub2015}
L.~G.~S. Jeub, P.~Balachandran, M.~A. Porter, P.~J. Mucha, and M.~W. Mahoney.
\newblock Think locally, act locally: detection of small, medium-sized, and
  large communities in large networks.
\newblock \emph{Physical Review E}, 91\penalty0 (1):\penalty0 012821, 2015.

\bibitem[Karrer and Newman(2011)]{degreeCorrect}
B.~Karrer and M.~E.~J. Newman.
\newblock Stochastic blockmodels and community structure in networks.
\newblock \emph{Physical Review E}, 83\penalty0 (1):\penalty0 016107, 2011.

\bibitem[Kernighan and Lin(1970)]{Kernighan1970}
B.~W. Kernighan and S.~Lin.
\newblock An efficient heuristic procedure for partitioning graphs.
\newblock \emph{Bell System Technical Journal}, 49\penalty0 (2):\penalty0
  291--307, 1970.

\bibitem[Kivel{\"a} et~al.(2014)Kivel{\"a}, Arenas, Barthelemy, Gleeson,
  Moreno, and Porter]{kivela2014}
M.~Kivel{\"a}, A.~Arenas, M.~Barthelemy, J.~P. Gleeson, Y.~Moreno, and M.~A.
  Porter.
\newblock Multilayer networks.
\newblock \emph{Journal of Complex Networks}, 2\penalty0 (3):\penalty0
  203--271, 2014.

\bibitem[Larremore et~al.(2013)Larremore, Clauset, and
  Buckee]{larremore2013network}
D.~B. Larremore, A.~Clauset, and C.~O. Buckee.
\newblock A network approach to analyzing highly recombinant malaria parasite
  genes.
\newblock \emph{PLoS Computational Biology}, 9\penalty0 (10):\penalty0
  e1003268, 2013.

\bibitem[Lee et~al.(2017)Lee, Magallanes, and Porter]{lee2017}
S.~H. Lee, J.~M. Magallanes, and M.~A. Porter.
\newblock Time-dependent community structure in legislation cosponsorship
  networks in the congress of the republic of peru.
\newblock \emph{Journal of Complex Networks}, 5\penalty0 (1):\penalty0
  127--144, 2017.

\bibitem[Levorato and Frota(2016)]{levorato2016}
M.~Levorato and Y.~Frota.
\newblock Brazilian congress structural balance analysis.
\newblock \emph{arXiv:1609.00767}, 2016.

\bibitem[Li et~al.(2008)Li, Leyva, Almendral, Sendi{\~n}a-Nadal, Buld{\'u},
  Havlin, and Boccaletti]{Li2008}
D.~Li, I.~Leyva, J.~A. Almendral, I.~Sendi{\~n}a-Nadal, J.~M. Buld{\'u},
  S.~Havlin, and S.~Boccaletti.
\newblock Synchronization interfaces and overlapping communities in complex
  networks.
\newblock \emph{Physical Review Letters}, 101\penalty0 (16):\penalty0 168701,
  2008.

\bibitem[Macon et~al.(2012)Macon, Mucha, and Porter]{macon2012}
K.~T. Macon, P.~J. Mucha, and M.~A. Porter.
\newblock Community structure in the united nations general assembly.
\newblock \emph{Physica A}, 391\penalty0 (1--2):\penalty0 343--361, 2012.

\bibitem[Mahoney et~al.(2012)Mahoney, Orecchia, and Vishnoi]{Mahoney2012}
M.~W. Mahoney, L.~Orecchia, and N.~K. Vishnoi.
\newblock A local spectral method for graphs: With applications to improving
  graph partitions and exploring data graphs locally.
\newblock \emph{Journal of Machine Learning Research}, 13\penalty0
  (1):\penalty0 2339--2365, 2012.

\bibitem[McPherson et~al.(2001)McPherson, Smith-Lovin, and
  Cook]{mcpherson2001birds}
M.~McPherson, L.~Smith-Lovin, and J.~M. Cook.
\newblock Birds of a feather: Homophily in social networks.
\newblock \emph{Annual Review of Sociology}, 27:\penalty0 415--444, 2001.

\bibitem[Meil\v{a}(2005)]{VI}
M.~Meil\v{a}.
\newblock Comparing clusterings: An axiomatic view.
\newblock In \emph{Proceedings of the 22Nd International Conference on Machine
  Learning}, ICML '05, pages 577--584, New York, NY, USA, 2005. ACM.
\newblock ISBN 1-59593-180-5.

\bibitem[Melnik et~al.(2013)Melnik, Ward, Gleeson, and Porter]{Melnik2013}
S.~Melnik, J.~A. Ward, J.~P. Gleeson, and M.~A. Porter.
\newblock Multi-stage complex contagions.
\newblock \emph{Chaos}, 23\penalty0 (1):\penalty0 013124, 2013.

\bibitem[Melnik et~al.(2014)Melnik, Porter, Mucha, and
  Gleeson]{melnik2014dynamics}
S.~Melnik, M.~A. Porter, P.~J. Mucha, and J.~P. Gleeson.
\newblock Dynamics on modular networks with heterogeneous correlations.
\newblock \emph{Chaos}, 24\penalty0 (2):\penalty0 023106, 2014.

\bibitem[Moody and Mucha(2013)]{moody2013}
J.~Moody and P.~J. Mucha.
\newblock Portrait of political party polarization.
\newblock \emph{Network Science}, 1\penalty0 (01):\penalty0 119--121, 2013.

\bibitem[Moody and White(2003)]{Moody2003}
J.~Moody and D.~R. White.
\newblock Structural cohesion and embeddedness: A hierarchical concept of
  social groups.
\newblock \emph{American Sociological Review}, 68\penalty0 (1):\penalty0
  103--127, 2003.

\bibitem[Mucha and Porter(2010)]{muchaporter2010}
P.~J. Mucha and M.~A. Porter.
\newblock Communities in multislice voting networks.
\newblock \emph{Chaos}, 20\penalty0 (4):\penalty0 041108, 2010.

\bibitem[Mucha et~al.(2010)Mucha, Richardson, Macon, Porter, and
  Onnela]{mucha2010community}
P.~J. Mucha, T.~Richardson, K.~Macon, M.~A. Porter, and J.-P. Onnela.
\newblock Community structure in time-dependent, multiscale, and multiplex
  networks.
\newblock \emph{Science}, 328\penalty0 (5980):\penalty0 876--878, 2010.

\bibitem[Nadakuditi and Newman(2012)]{nadakuditi2012}
R.~R. Nadakuditi and M.~E.~J. Newman.
\newblock Graph spectra and the detectability of community structure in
  networks.
\newblock \emph{Physical Review Letters}, 108:\penalty0 188701, 2012.

\bibitem[Newman and Clauset(2016)]{newman2016structure}
M.~E.~J. Newman and A.~Clauset.
\newblock Structure and inference in annotated networks.
\newblock \emph{Nature Communications}, 7:\penalty0 11863, 2016.

\bibitem[Newman(2002)]{assort}
M.~E.~J. Newman.
\newblock Assortative mixing in networks.
\newblock \emph{Physical Review Letters}, 89\penalty0 (20):\penalty0 208701,
  2002.

\bibitem[Newman(2006)]{Newman2006}
M.~E.~J. Newman.
\newblock Finding community structure in networks using the eigenvectors of
  matrices.
\newblock \emph{Physical Review E}, 74\penalty0 (3), 2006.

\bibitem[Newman and Girvan(2004)]{newmangirvan}
M.~E.~J. Newman and M.~Girvan.
\newblock Finding and evaluating community structure in networks.
\newblock \emph{Physical Review E}, 69:\penalty0 026113, 2004.

\bibitem[Onnela et~al.(2007)Onnela, Saram{\"a}ki, Hyv{\"o}nen, Szab{\'o},
  Lazer, Kaski, Kert{\'e}sz, and Barab{\'a}si]{onnela2007structure}
J.-P. Onnela, J.~Saram{\"a}ki, J.~Hyv{\"o}nen, G.~Szab{\'o}, D.~Lazer,
  K.~Kaski, J.~Kert{\'e}sz, and A.-L. Barab{\'a}si.
\newblock Structure and tie strengths in mobile communication networks.
\newblock \emph{Proceedings of the National Academy of Sciences}, 104\penalty0
  (18):\penalty0 7332--7336, 2007.

\bibitem[O'Sullivan et~al.(2015)O'Sullivan, O'Keeffe, Fennell, and
  Gleeson]{o2015mathematical}
D.~J. O'Sullivan, G.~J. O'Keeffe, P.~G. Fennell, and J.~P. Gleeson.
\newblock Mathematical modeling of complex contagion on clustered networks.
\newblock \emph{Frontiers in Physics}, 3:\penalty0 71, 2015.

\bibitem[Peel et~al.(2017)Peel, Larremore, and Clauset]{peel2016ground}
L.~Peel, D.~B. Larremore, and A.~Clauset.
\newblock The ground truth about metadata and community detection in networks.
\newblock \emph{Science Advances}, 3\penalty0 (5):\penalty0 E1602548, 2017.

\bibitem[Peixoto(2013)]{Peixoto2013}
T.~P. Peixoto.
\newblock Parsimonious module inference in large networks.
\newblock \emph{Physical Review Letters}, 110\penalty0 (14):\penalty0 148701,
  2013.

\bibitem[Peixoto(2014)]{Peixoto2014}
T.~P. Peixoto.
\newblock Hierarchical block structures and high-resolution model selection in
  large networks.
\newblock \emph{Physical Review X}, 4\penalty0 (1), 2014.

\bibitem[Pons and Latapy(2005)]{Pons2005}
P.~Pons and M.~Latapy.
\newblock \emph{Computing Communities in Large Networks Using Random Walks},
  pages 284--293.
\newblock Springer Berlin Heidelberg, Berlin, Heidelberg, 2005.
\newblock ISBN 978-3-540-32085-2.

\bibitem[Porter et~al.(2005)Porter, Mucha, Newman, and
  Warmbrand]{Porter_Mucha_Newman_Warmbrand_2005}
M.~A. Porter, P.~J. Mucha, M.~E.~J. Newman, and C.~M. Warmbrand.
\newblock A network analysis of committees in the {US House of
  Representatives}.
\newblock \emph{Proceedings of the National Academy of Sciences}, 102\penalty0
  (20):\penalty0 7057---7062, 2005.

\bibitem[Porter et~al.(2006)Porter, Friend, Mucha, and
  Newman]{Porter_Friend_Mucha_Newman_2006}
M.~A. Porter, A.~J. Friend, P.~J. Mucha, and M.~E.~J. Newman.
\newblock Community structure in the {U.S. House of Representatives}.
\newblock \emph{Chaos}, 16\penalty0 (4):\penalty0 041106, 2006.

\bibitem[Porter et~al.(2007)Porter, Mucha, Newman, and
  Friend]{Porter_Mucha_Newman_Friend_2007}
M.~A. Porter, P.~J. Mucha, M.~E.~J. Newman, and A.~J. Friend.
\newblock Community structure in the {United States House of Representatives}.
\newblock \emph{Physica A}, 386\penalty0 (1):\penalty0 414--438, 2007.

\bibitem[Porter et~al.(2009)Porter, Onnela, and Mucha]{porter2009}
M.~A. Porter, J.-P. Onnela, and P.~J. Mucha.
\newblock Communities in networks.
\newblock \emph{Notices of the AMS}, 56\penalty0 (9):\penalty0 1082--1097,
  1164--1166, 2009.

\bibitem[Power et~al.(2011)Power, Cohen, Nelson, Wig, Barnes, Church, Vogel,
  Laumann, Miezin, Schlaggar, and Petersen]{Power2011}
J.~D. Power, A.~L. Cohen, S.~M. Nelson, G.~S. Wig, K.~A. Barnes, J.~A. Church,
  A.~C. Vogel, T.~O. Laumann, F.~M. Miezin, B.~L. Schlaggar, and S.~E.
  Petersen.
\newblock Functional network organization of the human brain.
\newblock \emph{Neuron}, 72\penalty0 (4):\penalty0 665--678, 2011.

\bibitem[Reichardt and Bornholdt(2004)]{Reichardt2004}
J.~Reichardt and S.~Bornholdt.
\newblock Detecting fuzzy community structures in complex networks with a potts
  model.
\newblock \emph{Physical Review Letters}, 93\penalty0 (21):\penalty0 218701,
  2004.

\bibitem[Reichardt and Bornholdt(2006)]{Reichardt_Bornholdt_2006}
J.~Reichardt and S.~Bornholdt.
\newblock Statistical mechanics of community detection.
\newblock \emph{Physical Review E}, 74\penalty0 (1):\penalty0 016110, 2006.

\bibitem[Rosvall and Bergstrom(2008)]{rosvall2008maps}
M.~Rosvall and C.~T. Bergstrom.
\newblock Maps of random walks on complex networks reveal community structure.
\newblock \emph{Proceedings of the National Academy of Sciences}, 105\penalty0
  (4):\penalty0 1118--1123, 2008.

\bibitem[Rubenstein et~al.(2015)Rubenstein, Sundaresan, Fischhoff,
  Tantipathananandh, and Berger-Wolf]{Rubenstein2015}
D.~I. Rubenstein, S.~R. Sundaresan, I.~R. Fischhoff, C.~Tantipathananandh, and
  T.~Y. Berger-Wolf.
\newblock Similar but different: Dynamic social network analysis highlights
  fundamental differences between the fission-fusion societies of two equid
  species, the onager and grevy's zebra.
\newblock \emph{PLOS ONE}, 10\penalty0 (10):\penalty0 1--21, 10 2015.
\newblock \doi{10.1371/journal.pone.0138645}.
\newblock URL \url{https://doi.org/10.1371/journal.pone.0138645}.

\bibitem[Rubinov and Sporns(2010)]{Rubinov2010}
M.~Rubinov and O.~Sporns.
\newblock Complex network measures of brain connectivity: uses and
  interpretations.
\newblock \emph{NeuroImage}, 52\penalty0 (3):\penalty0 1059--1069, 2010.

\bibitem[Schaub et~al.(2017)Schaub, Delvenne, Rosvall, and
  Lambiotte]{Schaub2017}
M.~T. Schaub, J.-C. Delvenne, M.~Rosvall, and R.~Lambiotte.
\newblock The many facets of community detection in complex networks.
\newblock \emph{Applied Network Science}, 2\penalty0 (1):\penalty0 4, 2017.

\bibitem[Shalizi and Thomas(2011)]{shalizi2011homophily}
C.~R. Shalizi and A.~C. Thomas.
\newblock Homophily and contagion are generically confounded in observational
  social network studies.
\newblock \emph{Sociological Methods \& Research}, 40\penalty0 (2):\penalty0
  211--239, 2011.

\bibitem[Skardal and Restrepo(2012)]{skardal2012hierarchical}
P.~S. Skardal and J.~G. Restrepo.
\newblock Hierarchical synchrony of phase oscillators in modular networks.
\newblock \emph{Physical Review E}, 85\penalty0 (1):\penalty0 016208, 2012.

\bibitem[Snijders and Nowicki(1997)]{Snijders1997}
A.~T. Snijders and K.~Nowicki.
\newblock Estimation and prediction for stochastic blockmodels for graphs with
  latent block structure.
\newblock \emph{Journal of Classification}, 14\penalty0 (1):\penalty0 75--100,
  1997.

\bibitem[Spirin and Mirny(2003)]{Spirin2003}
V.~Spirin and L.~A. Mirny.
\newblock Protein complexes and functional modules in molecular networks.
\newblock \emph{Proceedings of the National Academy of Sciences}, 100\penalty0
  (21):\penalty0 12123--12128, 2003.

\bibitem[Sporns(2013)]{Sporns2013}
O.~Sporns.
\newblock Structure and function of complex brain networks.
\newblock \emph{Dialogues in Clinical Neuroscience}, 15\penalty0 (3):\penalty0
  247--262, 2013.

\bibitem[Sporns and Betzel(2016)]{Sporns2016}
O.~Sporns and R.~F. Betzel.
\newblock Modular brain networks.
\newblock \emph{Annual Review of Psychology}, 67:\penalty0 613--640, 2016.

\bibitem[Taylor et~al.(2015)Taylor, Klimm, Harrington, Kram{\'a}r, Mischaikow,
  Porter, and Mucha]{taylor2015topological}
D.~Taylor, F.~Klimm, H.~A. Harrington, M.~Kram{\'a}r, K.~Mischaikow, M.~A.
  Porter, and P.~J. Mucha.
\newblock Topological data analysis of contagion maps for examining spreading
  processes on networks.
\newblock \emph{Nature Communications}, 6:\penalty0 7723, 2015.

\bibitem[Traud et~al.(2011)Traud, Kelsic, Mucha, and Porter]{Traud2011}
A.~L. Traud, E.~D. Kelsic, P.~J. Mucha, and M.~A. Porter.
\newblock Comparing community structure to characteristics in online collegiate
  social networks.
\newblock \emph{SIAM Review}, 53\penalty0 (3):\penalty0 526--543, 2011.

\bibitem[Ugander et~al.(2012)Ugander, Backstrom, Marlow, and
  Kleinberg]{ugander2012structural}
J.~Ugander, L.~Backstrom, C.~Marlow, and J.~Kleinberg.
\newblock Structural diversity in social contagion.
\newblock \emph{Proceedings of the National Academy of Sciences}, 109\penalty0
  (16):\penalty0 5962--5966, 2012.

\bibitem[Waugh et~al.(2009)Waugh, Pei, Fowler, Mucha, and Porter]{waugh2009}
A.~S. Waugh, L.~Pei, J.~H. Fowler, P.~J. Mucha, and M.~A. Porter.
\newblock Party polarization in {Congress}: A network science approach.
\newblock \emph{arXiv:0907.3509}, 2009.

\bibitem[Weng et~al.(2013)Weng, Menczer, and Ahn]{weng2013virality}
L.~Weng, F.~Menczer, and Y.-Y. Ahn.
\newblock Virality prediction and community structure in social networks.
\newblock \emph{Scientific Reports}, 3:\penalty0 2522, 2013.

\bibitem[White et~al.(1976)White, Boorman, and Breiger]{white1976}
H.~C. White, S.~A. Boorman, and R.~L. Breiger.
\newblock Social structure from multiple networks. {I. Blockmodels} of roles
  and positions.
\newblock \emph{American Journal of Sociology}, 81\penalty0 (4):\penalty0
  730--780, 1976.

\bibitem[Wikipedia(2017{\natexlab{a}})]{CivilRightsAct1957}
Wikipedia.
\newblock Civil rights act of 1957, 2017{\natexlab{a}}.
\newblock URL
  \url{http://en.wikipedia.org/w/index.php?title=Civil_Rights_Act_of_1957}.
\newblock Page Version ID: 761682759, accessed March 7, 2017.

\bibitem[Wikipedia(2017{\natexlab{b}})]{PoliticalParties}
Wikipedia.
\newblock Political parties in the united states, 2017{\natexlab{b}}.
\newblock URL
  \url{https://en.wikipedia.org/w/index.php?title=Political_parties_in_the_United_States}.
\newblock Page Version ID: 768751459, accessed March 8, 2017.

\bibitem[Wu(1982)]{Wu1982}
F.~Y. Wu.
\newblock The {Potts} model.
\newblock \emph{Reviews of Modern Physics}, 54\penalty0 (1):\penalty0 235--268,
  1982.

\bibitem[Yang et~al.(2013)Yang, McAuley, and Leskovec]{yang2013community}
J.~Yang, J.~McAuley, and J.~Leskovec.
\newblock Community detection in networks with node attributes.
\newblock In \emph{2013 IEEE 13th International Conference on Data Mining},
  pages 1151--1156. IEEE, 2013.

\bibitem[Zachary(1977)]{Zachary1977}
W.~W. Zachary.
\newblock An information flow model for conflict and fission in small groups.
\newblock \emph{Journal of Anthropological Research}, 33\penalty0 (4):\penalty0
  452--473, 1977.

\bibitem[Zhang et~al.(2007)Zhang, Friend, Traud, Porter, Fowler, and
  Mucha]{Zhang_Friend_Traud_Porter_Fowler_Mucha_2007}
Y.~Zhang, A.~J. Friend, A.~L. Traud, M.~A. Porter, J.~H. Fowler, and P.~J.
  Mucha.
\newblock Community structure in {Congressional} cosponsorship networks.
\newblock \emph{Physica A}, 387\penalty0 (7):\penalty0 1705--1712, 2007.

\bibitem[Zhou(2003)]{Zhou2003}
H.~Zhou.
\newblock Network landscape from a {Brownian} particle's perspective.
\newblock \emph{Physical Review E}, 67\penalty0 (4):\penalty0 041908, 2003.

\end{thebibliography}

\end{document}